\documentclass[aoas,nameyear,seceqn,dvips]{arximspdf}
\usepackage{dcolumn}
\usepackage{graphicx}

\doi{10.1214/10-AOAS374}
\volume{5}
\issue{1}
\pubyear{2011}
\firstpage{486}
\lastpage{522}

\makeatletter

  \let\sv@tabnotetext\tabnotetext
  \let\sv@tabnotemark@fmt\tabnotemark@fmt
   \long\def\legend#1{{\let\tabnote@indent\leavevmode\sv@tabnotetext[]{}{#1}}}

\renewcommand{\epsilon}{\varepsilon}
\newcolumntype{d}[1]{D{.}{.}{#1}}
\newcolumntype{k}[1]{D{.}{}{#1}}
\newproclaim{remark}{Remark}
\makeatother

\begin{document}
\begin{frontmatter}

\title{Orthogonal simple component analysis: A new, exploratory approach}
\runtitle{Orthogonal simple component analysis}

\begin{aug}
\author[A]{\fnms{Karim} \snm{Anaya-Izquierdo}\corref{}\ead[label=e1]{k.anaya@open.ac.uk}},
\author[A]{\fnms{Frank} \snm{Critchley}\ead[label=e2]{f.critchley@open.ac.uk}}
\and
\author[A]{\fnms{Karen} \snm{Vines}\ead[label=e3]{s.k.vines@open.ac.uk}}

\runauthor{K. Anaya-Izquierdo, F. Critchley and K. Vines}

\affiliation{The Open University}

\address[A]{Department of Mathematics and Statistics\\
The Open University\\
Walton Hall\\
Milton Keynes \\
MK7 6AA \\
United Kingdom\\
\printead{e1}\\
\phantom{E-mail:\ }\printead*{e2} \\
\phantom{E-mail:\ }\printead*{e3}} 

\end{aug}

\received{\smonth{3} \syear{2010}}
\revised{\smonth{6} \syear{2010}}

\begin{abstract}
Combining principles with pragmatism, a new approach and accompanying
algorithm are presented to a longstanding problem in applied
statistics: the interpretation of principal components. Following
Rousson and Gasser~[\textbf{53} (2004) 539--555]\vspace*{4pt}
\begin{center}
\begin{tabular}{@{}p{250pt}@{}}
the ultimate goal is not to propose a method that leads automatically
to a unique solution, but rather to develop tools for assisting the
user in his or her choice of an interpretable solution.
\end{tabular}
\end{center}
\vspace*{4pt}
Accordingly, our approach is essentially \textit{exploratory}.
Calling a vector `simple' if it has small integer elements, it poses
the open question:\vspace*{4pt}
\begin{center}
\begin{tabular}{@{}p{250pt}@{}}
What sets of simply interpretable orthogonal axes---if any---are
angle-close to the principal components of interest?
\end{tabular}
\end{center}
\vspace*{4pt}
its answer being presented in summary form as an automated
visual display of the solutions found, ordered in terms of overall
measures of simplicity, accuracy and star quality, from which the user
may choose. Here, `star quality' refers to striking overall patterns in
the sets of axes found, deserving to be especially drawn to the user's
attention precisely because they have emerged from the data, rather
than being imposed on it by (implicitly) adopting a model. Indeed,
other things being equal, explicit models can be checked by seeing if
their fits occur in our exploratory analysis, as we illustrate.
Requiring orthogonality, attractive visualization and dimension
reduction features of principal component analysis are retained.

Exact implementation of this principled approach is shown to provide an
exhaustive set of solutions, but is combinatorially hard.
Pragmatically, we provide an efficient, approximate algorithm.
Throughout, worked examples show how this new tool adds to the applied
statistician's armoury, effectively combining simplicity, retention of
optimality and computational efficiency, while complementing existing
methods. Examples are also given where simple structure in the
population principal components is recovered using only information
from the sample. Further developments are briefly indicated.
\end{abstract}


\end{frontmatter}

\section{Introduction and overview}\label{sec:intro}

Principal components are linear combinations of a set of, say, $p$
commensurable variables with coefficients (`loadings') given by
eigenvectors of their covariance or correlation matrix $\mathbf{S}$. As
such, they simultaneously enjoy many optimal properties: see,
for
example, \citet{Joll2002}, Chapters 2 and 3. However, to be useful in
practice, such components often need interpretation in the context of
the data studied. Unfortunately, optimality is no guarantee of
interpretability. Accordingly, principal components may possess optimal
theoretical properties, but be of limited practical interest. This
motivates replacing them by components which are more interpretable by
virtue of being `simpler' in some sense, albeit at the expense of some
degree of optimality.

We begin with a brief overview of existing approaches to this problem,
further details being available in the references cited.

\subsection{Existing approaches}

In a broad sense, simplicity means the appearance of nice structures in
the loadings matrix $\mathbf{Q=(q}_{1}|\cdots|\mathbf{q}_{k}\mathbf{)}$
which contains the $k\leq p$ eigenvectors of interest. Often, the
scientist in charge of the study would like to see if there are
clear-cut patterns reflected in $\mathbf{Q}$ which help him or her to
better understand the meaning of the components
$\mathbf{q}_{r}^{\top}\mathbf{x}$ $(r=1,\ldots,k)$ which it generates.
Examples of nice structures include the presence of simple weighted
averages, contrasts, groups of variables and sparseness. However
defined, simplicity inevitably implies some loss of optimality and it
is the scientist in charge of the study who needs to calibrate the
trade-off between simplicity and optimality, as we further comment in
Section \ref{sec:inter}.

The oldest approach to simplifying principal components is rotation,
exploiting the fact that---as with principal component analysis itself---rotation
of the $p$ original axes (one for each variable) defines
new orthogonal coordinate axes on which the data can be displayed while
total variance is preserved. This provides attractive visualization and
dimension reduction features. In particular, there being no double
counting of total variance, the user can identify and plot the data on
just those axes making the largest or smallest contributions to it,
depending on the focus of scientific interest---explaining variability
or exploring potential scientific laws (near constant linear relations
among the variables).

Only rotation methods are guaranteed to provide new axes which are
orthogonal. Nonrotation methods in general lack the attractive features
noted above, joint visualization of components being impeded by
nonorthogonality of axes and dimension reduction by loss of the
additive decomposition of total variance.

Overall, the rotation approach to simplification seeks more
interpretable, orthogonal axes while retaining as much optimality as
possible. See, for example, Chapter 11 of \citet{Joll2002}, which
provides an excellent overall review of simplification of principal
components as of 2002. More recently, assuming normality,
\citet{Park2005} has proposed a penalized profile likelihood method,
using varimax as the penalty function, which favors rotation of
ill-defined components (those whose eigenvalues are close). However, in
all these methods, the loadings involved are usually real numbers,
which means that interpretation can still be difficult.

Another approach to simplification is to target sparsity. The presence
of many zeroes in $\mathbf{Q}$ can be useful for interpretation, for
example, when dealing with many variables. See, for example,
\citet{DaspElghJordLanc2005},
\citet{Farc2006}, \citet{ChipGu2005} and the references therein. One
class of methods which targets sparseness is that based on the Least
Absolute Shrinkage Selection Operator (LASSO). See, for example, the
papers by \citet{JollTren2007}, \citet{ZouHastTibs2006},
\citet{SjosStegLars2006} and \citet{JollTrenUddi2003}. Although
most of these methods lead to orthogonal simplified components,
combined with the presence of exact zero loadings, the remaining
loadings are still real numbers, again impeding interpretation.

Other approaches simplify by imposing specific structures on the
original data matrix $\mathbf{X}$ and can be seen as constrained
singular value decompositions. For example, the semidiscrete
decomposition (SDD) approach of \citet{KoldOlea1998} restricts the
loadings to lie in $\{-1,0,1\}$. Again, the nonnegative matrix
factorization (NMF) approach of \citet{LeeSeung1999} requires the
original variables to be nonnegative, decomposing $\mathbf{X}$ into
two nonnegative factor matrices. More recently, plaid models [see, for
example, \citet{LazzeOwen2002}] impose various block structures on
$\mathbf{X}$ which are useful for interpretation in gene expression
microarray data. However, this class of methods does not require
orthogonality of the simplified components, with the potential loss of
attractive features noted above.

A more explicitly modeling approach to simplification has recently been
suggested in \citet{RousGass2004}. Intrinsically restricted to
principal component analysis of a correlation matrix, it assumes a
particular pattern in the eigenstructure of that matrix in which groups
and contrasts of variables are forced to appear. Although not always
appropriate, it is when all variables are positively correlated, the
first eigenvector being then a weighted average of the variables and,
consequently, the remaining eigenvectors being basically contrasts. The
loadings obtained are all proportional to integers, aiding
interpretation. However, the components obtained need not be
orthogonal, again with the potential drawbacks noted above.

The approaches in \citet{Haus1982}, \citet{SunL2006} and
\citet{Vine2000} are similar to the one presented here, in the sense
that all three give orthogonal components with loading vectors
proportional to integers. Hausman's method only allows the loadings to
take the values $-1$, $0$ or $1$ and so is not always able to find a
complete set of orthogonal vectors. In contrast, Vines' method produces
loading vectors that are proportional to integers via a sequence of
pairwise `simplicity-preserving' transformations which ensure that
orthogonality is maintained. However, although always proportional to
integers, the size of the integers is not bounded and may at times be
very large. A fuller discussion of the method, and its properties, can
be found in \citet{SunL2006}.

\subsection{Interpretability}\label{sec:inter}

With others, we note that interpretability is neither guaranteed nor
amenable to precise mathematical formulation, this latter being
evidenced by the variety both between and within methods reviewed
above.

These remarks have two key methodological consequences. First, whereas
simplification can help in the vital step of interpretation, we do not
expect any method to lead to interpretable results in all cases. And
second, rather than attempt to find a unique optimal simplification in
any predefined sense---in particular, rather than attempt to
completely automate the trade-off between simplicity and optimality---they
provide motivation for adopting an essentially exploratory
approach which systematically produces an ordered range of solutions,
from which the user can choose one or more preferred solutions.

Factors that can guide this choice include the following: (a) the
criteria on which a method is based, (b) subject matter considerations,
particular to the context of the data set under analysis, and (c)
suboptimality with respect to exact principal component analysis,
including loss of explanatory power---or of focus on potential
scientific laws---and correlation. Only principal component analysis
itself can give orthogonal loadings and uncorrelated components, and so
any other rotation method will always show some degree of correlation.

\subsection{Overview of a new approach}

Beginning with a synoptic account, we give here an overview of our new,
exploratory approach. Requiring orthogonality, it is based on three
primary criteria: simplicity, angle-accuracy and `star quality.'

\subsubsection{A synoptic account}

Retaining the attractive visualization and dimension reduction features
noted above, the approach to be presented is based on rotation to axes
which are `simple' in the sense---\textit{adopted henceforth}---that
each is defined by small integer loadings. It combines principles with
pragmatism, complementing those already available. Following
\citet{RousGass2004},

\begin{quote}
the ultimate goal is not to propose a method that
leads automatically to a unique solution, but rather to develop tools
for assisting the user in his or her choice of an interpretable
solution.
\end{quote}
Accordingly, our approach is essentially
\textit{exploratory}, posing the open question:

\begin{quote}
What sets of simply interpretable orthogonal axes---if
any---are angle-close to the principal components of
interest?
\end{quote}
its answer being presented in summary form as an automated
visual display of the solutions found, ordered in terms of overall
measures of simplicity, angle-accuracy and star quality, from which the
user may choose.

Here, `star quality' refers to striking overall patterns in the sets of
axes found, deserving to be especially drawn to the user's attention
precisely because they have emerged from the data, rather than been
imposed on it by (implicitly) adopting a model. Indeed, other things
being equal, explicit models can be checked by seeing if their fits
occur in our essentially exploratory analysis, as we illustrate.

Our approach treats the components of interest equally, reflecting
equal scientific interest in them. Along with later worked examples,
the one that follows illustrates the appropriateness of adopting this
principle. Adaptations of our methodology to other scientific contexts---notably,
to those where interest focuses \textit{exclusively} on
explaining variability---are noted in Section \ref{sec:discuss}.

Again, our approach trades angle-accuracy off against simplicity, with
a bias toward the latter. Its exact implementation provides an
exhaustive set of solutions but can be prohibitively hard, the solution
space having combinatorial complexity which grows with $p$, $k$ and
$N^{\ast}$, the maximum size of integer allowed. However, the nature of
our approach allows efficient exploration of this vast space without
restriction to any of its particular subsets, such as those determined
by modeling assumptions. Pragmatically, we are able to provide an
efficient, approximate algorithm for this computationally challenging
problem.

\subsubsection{A worked example: Blood flow data}\label{sec:ri}

\begin{figure}[b]

\includegraphics{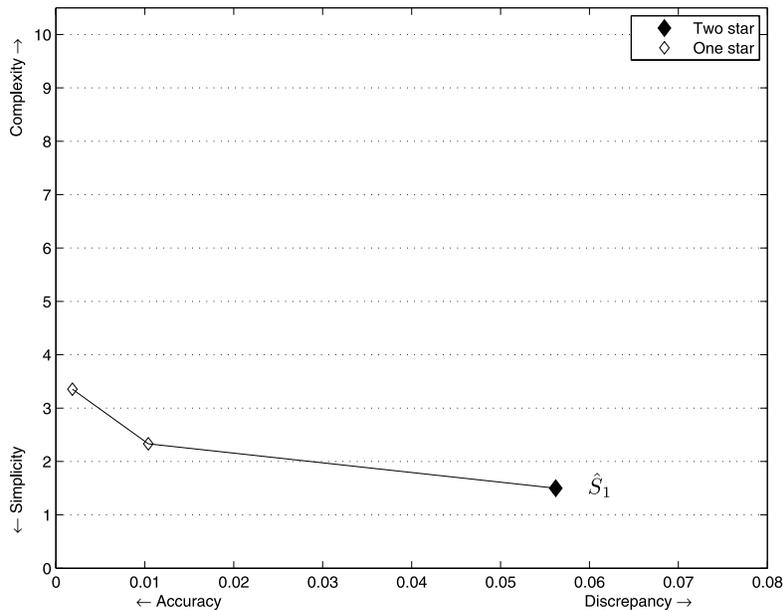}

\caption{A graphical summary of the solutions obtained for the blood
data by our approach.} \label{fig:ri}
\end{figure}

A worked example illustrates this new approach. Figure \ref{fig:ri},
whose construction and terms are described in
Section \ref{sec:new.app}, summarizes its results on the covariance
matrix for four different measurements of an index of resistance to
flow in blood vessels [see the paper by \citet{ThomVineHarr1999}].
Here, $p=k=4$ and, as throughout the paper, we take the maximum integer
allowed  ($N^{\ast}$) to be  $9$, this corresponding to allowing only
single digit representations.

Three solutions are obtained and ordered as shown, none of them being
dominant in terms of both simplicity and accuracy. The user is referred
first to the one `two star' solution found, $\hat{S}_{1}$, also
obtained by \citet{Vine2000} and by the undeflated form of
\citet{ChipGu2005} [recall that \citet{RousGass2004} cannot be used
for covariance matrices]. This two star solution is also the simplest
one found in this case, details being shown in Table \ref{tb:ri.exact}
along with the original principal components.

\begin{table}
\caption{Principal component loadings for the blood flow data $\mathbf{q}%
_{1},\ldots,\mathbf{q}_{4}$ and the corresponding simplified loading
vectors $\hat{\mathbf{z}}_{1},\ldots,\hat{\mathbf{z}}_{4}$ for solution $\widehat{S}_{1}$}\label{tb:ri.exact}
\begin{tabular*}{\textwidth}{@{\extracolsep{\fill}}lcd{2.1}d{2.2}cd{2.2}cd{2.2}c@{}}
\hline
\textbf{Variable} & \multicolumn{1}{c}{$\bolds{\mathbf{q}_{1}}$} & \multicolumn{1}{c}{$\bolds{\hat{\mathbf{z}}_{1}}$}
& \multicolumn{1}{c}{$\bolds{\mathbf{q}_{2}}$} & \multicolumn{1}{c}{$\bolds{\hat{\mathbf{z}}_{2}}$} &
\multicolumn{1}{c}{$\bolds{\mathbf{q}_{3}}$} & \multicolumn{1}{c}{$\bolds{\hat{\mathbf{z}}_{3}}$} &
\multicolumn{1}{c}{$\bolds{\mathbf{q}_{4}}$} &
\multicolumn{1}{c@{}}{$\bolds{\hat{\mathbf{z}}_{4}}$}\\
\hline
Right doppler & 0.42 & 1 & -0.32 & $-$1\phantom{0,} & -0.58 & $-$1\phantom{0,} & -0.62 & $-$1\phantom{0,}\\
Left doppler & 0.43 & 1 & 0.30 & 1 & -0.55 & $-1$\phantom{0,} & 0.65 & 1\\
Right CVI & 0.55 & 1 & -0.65  & $-$1\phantom{0,} & 0.43 & 1 & 0.30 & 1\\
Left CVI & 0.58 & 1 & 0.63 & 1 & 0.42 & 1 & -0.31 & $-$1\phantom{0,}\\ [6pt]
Variance (\%) & 58.0\phantom{00} & \phantom{0}57.0 & 25.9 & 23.8 & 9.5 & 10.5 & 6.5 & \phantom{0}8.6\\
\hline
\end{tabular*}
\end{table}

The simplified loadings here have a very clear structure and are easier
to understand than the continuous ones, so much so, in fact, that it
looks like we have uncovered nature's design: a main effect, plus three
orthogonal contrasts. The simplified components being orthogonal, the
total variance is retained, being redistributed among the components so
as to enhance interpretability. In particular, there is just a little
loss in the variance explained by the first two components, while the
relatively small variability in the last two suggests possible
underlying regularities.

The user is referred next to the other, here `one star,' solutions,
starting with the simpler one. Having the same sign pattern, just
different weights, they have essentially the same overall
interpretation as $\hat{S}_{1}$. Successively gaining accuracy at the
cost of some simplicity, the simplified loading vectors and percentages
of variance explainedfor $\hat{S}_{2}$ and $\hat{S}_{3}$ are given
in Table \ref{tb:exams.s2}.

\begin{table}[b]\tablewidth=280pt
\tabcolsep=0pt
\caption{Integer representations of solutions $\hat{S}_{2}$ and
$\hat{S}_{3}$ for the blood flow data} \label{tb:exams.s2}
\begin{tabular*}{280pt}{@{\extracolsep{\fill}}lcccccccc@{}}
\hline
&\multicolumn{4}{c}{$\bolds{\hat{\bolds{S}}_{2}}$} \hspace*{6pt}&
\multicolumn{4}{c@{}}{$\bolds{\hat{\bolds{S}}_{3}}$}\\ [-7pt]
&\multicolumn{4}{c}{\hrulefill}\hspace*{6pt}&\multicolumn{4}{c@{}}{\hrulefill}\\
\textbf{Variable} & $\bolds{\hat{\mathbf{z}}_{1}}$ & $\bolds{\hat{\mathbf{z}}_{2}}$ & $\bolds{\hat{\mathbf{z}%
}_{3}}$ & $\bolds{\hat{\mathbf{z}}_{4}}$ \hspace*{6pt}& $\bolds{\hat{\mathbf{z}}_{1}}$ & $\bolds{\hat{\mathbf{z}%
}_{2}}$ & $\bolds{\hat{\mathbf{z}}_{3}}$ & $\bolds{\hat{\mathbf{z}}_{4}}$\\
\hline
Right doppler & 1 & $-$1\phantom{0,} & $-$1\phantom{0,} & $-$2\phantom{000} & 2 & $-$1\phantom{0,} & $-$3\phantom{0,} & $-$2\phantom{0,}\\
Left doppler & 1 & 1 & $-$1\phantom{0,} & 2\phantom{0,} & 2 & 1 & $-$3\phantom{0,} & 2\\
Right CVI & 1 & $-$2\phantom{0,} & 1 & 1\phantom{0,} & 3 & $-$2\phantom{0,} & 2 & 1\\
Left CVI & 1 & 2 & 1 & $-$1\phantom{000} & 3 & 2 & 2 & $-$1\phantom{0,}\\
Variance (\%)& 57.0 &\phantom{,}25.9 & \phantom{,}10.5 & 6.5
& 57.9 & \phantom{,}25.9 & \phantom{0,}9.7 & \phantom{0,}6.5\\
\hline
\end{tabular*}
\end{table}

Overall, as this and later examples will show, we can obtain good
approximations in the sense that only small integers are used, while
retaining closeness to the original components and exact orthogonality.
A distinctive feature of our exploratory approach is that the user is
provided with an ordered set of alternative views of the same data,
from which s/he may choose.

We move now to put some flesh on the bones of the synoptic account
above, noting first that intrinsic interest lies in eigenaxes, not
eigenvectors.

\subsubsection{Eigenvectors, eigenaxes and their approximation}\label{sec:evec}

Recall that interest centers on a $p\times k$ loadings matrix $\mathbf{Q=(q}%
_{1}|\cdots|\mathbf{q}_{k}\mathbf{)}$ containing the eigenvectors of
interest. Without loss,  these are normalized to unit length
($ \Vert \mathbf{q}_{r} \Vert =1,$ $r=1,\ldots,k$),
$\lambda_{1}>\cdots>\lambda_{k}$ being the corresponding eigenvalues.
The overall sign of each eigenvector is arbitrary. Rather, interest
really centers on
the ordered set of axes $\mathbf{\pm Q:=(\pm q}_{1}|\cdots|\mathbf{\pm q}%
_{k}\mathbf{)}$, where we identify any pair of nonzero opposed vectors
$\mathbf{\pm q}$ with the axis (line through the origin or
one-dimensional subspace) $\ell
(\mathbf{q}):=\{c\mathbf{q\dvtx-\infty}<c<\mathbf{\infty}\}$ containing
them.

The approach taken here treats the columns of $\mathbf{\pm Q}$ equally.
It retains their orthogonality while replacing each eigenaxis $\alpha
=\ell(\mathbf{q})$ by another one $\hat{\alpha}=\ell(\hat{\mathbf{z}%
})$, close to it in angle terms, which is `simple' in the sense that it
contains a nonzero vector $\hat{\mathbf{z}}$ with small integer
elements. There is no loss in taking the highest common factor of the
absolute values of the nonzero elements of $\hat{\mathbf{z}}$,
denoted $\operatorname{hcf}(|\hat {\mathbf{z}}|)$, to be $1$. For, if not, we can
divide each element of $\hat{\mathbf{z}}$ by it, without changing
$\ell(\hat{\mathbf{z}})$.

Overall then, $\mathbf{\pm Q}$ is approximated by $\mathbf{\pm}%
\widehat{\mathbf{Z}} :=(\pm\hat{\mathbf{z}}_{1}|\cdots|\mathbf{\pm
}\hat{\mathbf{z}}_{k}\mathbf{)}$ where  $\widehat{\mathbf{Z}}%
 :=(\hat{\mathbf{z}}_{1}|\cdots|\hat{\mathbf{z}}_{k}\mathbf{)}$
belongs to the set $\mathcal{Z}(p,k)$ of all $p\times k$ integer
matrices with nonzero, pairwise orthogonal columns in each of which the
absolute values of the nonzero elements are coprime, two members of
this set being \textit{axis-equivalent} if they differ, at most, in the
overall signs of their columns.

\subsubsection{Four maxims}\label{sec:maxims}

Our approach is driven by four maxims, adopted for specific
methodological reasons. Briefly, these are as follows.

(1) \textit{Integers aid interpretation.}
This maxim speaks for itself: we require linear combinations of
variables defined by simple vectors since they are typically much
easier to interpret than the principal components which they
approximate. Again, exact zeroes and simple averages appear naturally.

Approximating an eigenaxis $\alpha=l(\mathbf{q})$ by a simple axis
$\hat{\alpha}=l(\hat{\mathbf{z}})$ where $\operatorname{hcf}(|\hat{\mathbf{z}%
}|)=1$, we call $\hat{\mathbf{z}}$ an \textit{integer
representation} of $\hat{\alpha}$ and the maximum absolute value of
its elements the \textit{complexity} of $\hat{\mathbf{z}}$---interchangeably,
of $\hat{\alpha}$---denoting these complexities
by $\mathit{compl}(\hat{\mathbf{z}})\equiv \mathit{compl}(\hat{\alpha})$.

Other things being equal, we seek to keep the complexity of each
$\hat{\alpha}_{r}$ $(r=1,\ldots,k)$ as low as possible.

(2) \textit{Be angle accurate (for the $k$ eigenvectors of interest).}
By keeping each approximating vector angle-close to its exact
counterpart, we ensure that we do not lose potentially meaningful
individual eigenvectors and that overall optimality is maximally
retained. This is consistent with our principle of equal treatment of
all the eigenaxes of interest, while providing a natural, operational
measure of discrepancy, both for each axis separately and---it turns
out---overall.

More specifically, we measure the discrepancy with which a simple axis
$\hat{\alpha}=\mathbf{\pm}\hat{\mathbf{z}}$ approximates an
eigenaxis $\alpha=\mathbf{\pm q}$ by the acute angle
%
\begin{equation}\label{eq:dist}
d(\alpha,\hat{\alpha}):=\arccos \biggl( \frac{|\mathbf{q}^{\top}\hat
{\mathbf{z}}|}{\Vert\mathbf{q}\Vert\Vert\hat{\mathbf{z}}\Vert} \biggr)
\end{equation}
between them, this being a (geodesic) distance measure between axes.
Equivalently, for reporting purposes, we may use the accuracy measure
\[
\mathit{accu}(\alpha,\hat{\alpha}):=\cos(d(\alpha,\hat{\alpha})),
\]
this taking values in $[0,1]$.

It turns out that, when approximating each of a set of axes, the
greater the minimum angle-accuracy attained overall, the closer the
original and approximating sets are in terms of a natural measure of
distance (see Appendix~\ref{app:min.dist}).

\begin{figure}

\includegraphics{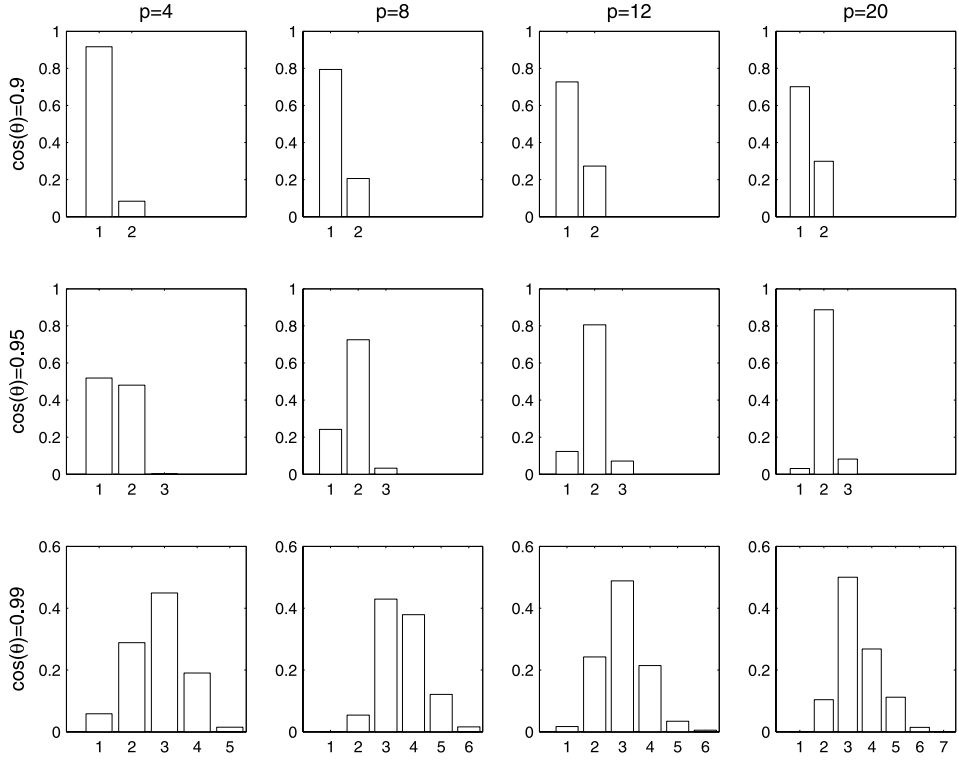}

\caption{Empirical distribution of minimum complexities $N_1(\theta)$
in simple approximations to $p$-dimensional space.} \label{fig:theta}
\end{figure}

(3) \textit{Be biased toward simplicity.}
It is always possible to approximate with reasonably high accuracy a
single $p$-dimensional axis $\ell(\mathbf{q})$ by a simple axis of low
complexity. Figure \ref{fig:theta} shows, for different values of $p$
and $\cos(\theta)$, the empirical distribution (based on 10,000
independent replications) of the minimum complexity $N_{1}(\theta)$
required for there to be a simple axis having accuracy greater than
$\cos(\theta)$ when $\ell(\mathbf{q})$ is sampled from the uniform
distribution over the set of all possible axes [see
\citet{FangLi1997}]. Clearly, without orthogonality restrictions,
accurate approximations of axes tend not to be very complex.

However, there is a clear trade-off between simplicity and accuracy:
highly accurate approximations usually have high complexity, making
interpretation more difficult. In general, we choose the simplest
possible axis that is accurate enough, this bias toward simplicity
being, in effect, a bias toward interpretability. In other words, in
case of conflict, we favor maxim 1 over maxim 2.

(4) \textit{Orthogonality brings benefits.}
Primarily, we choose orthogonality because it aids interpretation. Our
rotation approach enjoys the general visualization and dimension
reduction features recalled above. Although none of these additional
features is either targeted or imposed, sparsity, contrasts, simple
relations between components and groups of variables may all emerge as
a consequence of using orthogonality combined with integer
coefficients. Orthogonality is also useful at several stages of the
development, as we note.

\subsection{Organization and running example}\label{sec:org}

Section \ref{sec:new.app} develops these maxims into a methodology, the
running example below being used for illustration throughout. The
reader interested primarily in how this new approach performs may wish
to skip this development and go straight to
Section \ref{sec:exams.res}, where its results are summarized. Further
examples are given in Section \ref{sec:extra.ex}.
Section \ref{sec:discuss} gives a short discussion of complements and
extensions. Technical and computational details are given as
Appendices.

\begin{table}[b]\tablewidth=230pt
\caption{Principal component loadings for the exams data}\label{eigenvectors books data correlation}%
\tabcolsep=0pt
\begin{tabular*}{230pt}{@{\extracolsep{\fill}}lccccc@{}}
\hline
&$\bolds{\mathbf{q}_{1}}$ & $\bolds{\mathbf{q}_{2}}$ &
$\bolds{\mathbf{q}_{3}}$ & $\bolds{\mathbf{q}_{4}}$ & $\bolds{\mathbf{q}_{5}}$\\
\hline
Mechanics (closed) & 0.40 & $-$0.65 & \phantom{$-$}0.62 & \phantom{$-$}0.15 & $-$0.13\\
Vectors (closed) & 0.43 & $-$0.44 & $-$0.71 & $-$0.30 & $-$0.18\\
Algebra (open) & 0.50 & \phantom{$-$}0.13 & $-$0.04 & \phantom{$-$}0.11 & \phantom{$-$}0.85\\
Analysis (open) & 0.46 & \phantom{$-$}0.39 & $-$0.14 & \phantom{$-$}0.67 & $-$0.42\\
Statistics (open) & 0.40 & \phantom{$-$}0.47 & \phantom{$-$}0.31 & $-$0.66 & $-$0.23\\ [3pt]
Variance (\%)& 63.6\phantom{00} & 14.8\phantom{,} & \phantom{,}8.9 & \phantom{,}7.8 & \phantom{,}4.9\\
\hline
\end{tabular*}
\end{table}

The running example used is based on Table \ref{eigenvectors books data
correlation} which shows the unit length eigenvectors (rounded to 2
decimal places) of the sample correlation matrix for a data set
consisting of the scores achieved by $88$ students in $p=k=5$
tests, a combination of open- and closed-book exams [\citet{MardKentBibb1979}].
Thus, the first principal component is a weighted average of all the
different subject scores, while the other principal components can be
interpreted as contrasts. However, more detailed interpretation of the
principal components, particularly those other than the first, is not
easy.

Throughout, $\mathbb{Z}^{p}$ denotes the set of all $p\times1$ vectors
with integer elements---positive, negative or zero---and
$\mathbb{Z}^{(p)}$ the same set with the zero vector removed.
Replacing integers by real numbers, the corresponding sets are denoted
$\mathbb{R}^{p}$ and $\mathbb{R}^{(p)}$, respectively.

\section{A new approach} \label{sec:new.app}

\subsection{A sequential approach} \label{sec:seq.app}

Operationally, we approximate the $k$ eigenaxes of interest\
sequentially. The order in which we do this matters, for two principal
reasons: earlier approximations restrict the approximations available
for later eigenaxes and, hence, their maximum possible achievable
accuracy.

To illustrate these points consider, say, the `forwards' $1$ to $k$
order from high to low eigenvalue. When dealing with $\alpha_{1}$,
there are no orthogonality restrictions and we seek an approximation
$\hat{\alpha}_{1}$ to it in the set $\mathcal{M}_{1}$ of all simple
axes in $\mathbb{R}^{p}$. In contrast, for each $r\in\{2,\ldots,k\}$,
we seek an approximation $\hat {\alpha}_{r}$ to $\alpha_{r}$ within the
set $\mathcal{M}_{r}$ of all simple axes in $\mathbb{R}^{p}$ orthogonal
to each of $\hat{\alpha}_{1},\ldots ,\hat{\alpha}_{r-1}$.

Thus, for the exams data, $\mathcal{M}_{1}$ is the set of all axes
generated
by vectors in $\mathbb{Z}^{(5)}$ while, for example, taking $\hat{\mathbf{z}%
}_{1}=(1,1,1,1,1)^{\top}$, $\ell((1,1,0,-1,-1)^{\top})$ is a member of
$\mathcal{M}_{2}$, but $\ell((1,1,0,0,-1)^{\top})$ is not.

The second point is clear geometrically. The angle-closest axis to
$\alpha$ orthogonal to $\hat{\alpha}_{1},\ldots,\hat{\alpha}_{r-1}$ is
its projection onto the orthogonal complement of their span. This
restricts the maximum accuracy that can be achieved. For, if
$\mathbf{q}_{r}^{\perp}$ is the orthogonal projection of the unit
vector $\mathbf{q}_{r}$ onto the orthogonal complement of
$\operatorname{Span}\{\mathbf{z}_{1},\ldots,\mathbf{z}_{r-1}\}$, some
straightforward trigonometry shows that any approximation $\hat{\alpha}%
\in\mathcal{M}_{r}$ satisfies
\begin{eqnarray*}
\mathit{accu}(\alpha_{r},\hat{\alpha}) &=&\mathit{accu}(\alpha_{r},\ell(\mathbf{q}_{r}^{\bot
})) \mathit{accu}(\ell(\mathbf{q}_{r}^{\bot}),\hat{\alpha})\\
& =&\Vert\mathbf{q}_{r}^{\perp}\Vert \mathit{accu}(\ell(\mathbf{q}_{r}^{\perp}%
),\hat{\alpha}),%
\end{eqnarray*}
so that
$\mathit{accu}(\alpha_{r},\hat{\alpha})\leq\Vert\mathbf{q}_{r}^{\perp}\Vert$,
equality holding if and only if
$\hat{\alpha}=\ell(\mathbf{q}_{r}^{\perp})$\break [which requires
$\ell(\mathbf{q}_{r}^{\perp})$ to be simple]. Thus, over
$\mathcal{M}_{r}$, not every possible accuracy is achievable for
$\alpha_{r}$
($r>1$), although no such upper bound applies to
$\mathit{accu}(\ell(\mathbf{q}_{r}^{\perp}),\hat{\alpha})$.
\begin{table}
\caption{Integer representations for the examinations data with
$\theta=\pi /4$} \label{E2}
\begin{tabular*}{\textwidth}{@{\extracolsep{\fill}}lccccc@{}}
\hline
\textbf{Variable} & $\bolds{\hat{\mathbf{z}}_{1}(\theta)}$ &
$\bolds{\hat{\mathbf{z}}_{2}(\theta)}$ & $\bolds{\hat{\mathbf{z}}_{3}(\theta)}$ &
$\bolds{\hat{\mathbf{z}}_{4}(\theta)}$ & $\bolds{\hat{\mathbf{z}}_{5}(\theta)}$\\
\hline
Mechanics (closed) & 1\phantom{000} & 1\phantom{000} & 1\phantom{0000} & 0\phantom{00000}& 1\phantom{00}\\
Vectors (closed) & 1\phantom{000} & 1\phantom{000} & $-$1\phantom{00000,} & 0\phantom{00000} & 1\phantom{00}\\
Algebra (open) & 1\phantom{000} & 0\phantom{000} & 0\phantom{0000} & 0\phantom{00000} & $-$4\phantom{000,}\\
Analysis (open) & 1\phantom{000} & $-$1\phantom{0000,} & 0\phantom{0000} & 1\phantom{00000} & 1\phantom{00}\\
Statistics (open) & 1\phantom{000} & $-$1\phantom{0000,} & 0\phantom{0000} & $-$1\phantom{000000,} & 1\phantom{00}\\ [6pt]
Accuracy & 0.997 & 0.973 & 0.9375 & \textbf{0.937}\phantom{00} & \phantom{0}0.974\\ [3pt]
Max accuracy $\|\mathbf{q}^{\perp}_{r}\|$ & 1\phantom{000} & 0.999
& 0.99\phantom{00} & 0.95\phantom{000} & 0.97\\ [3pt]
Variance (\%)& 63.3\phantom{000} & 14.4\phantom{000} & 8.9\phantom{000} & 7.9\phantom{0000} &
5.5\phantom{0}\\
\hline
\end{tabular*}
\end{table}

For the exams data with $\hat{\alpha}_{1}=\ell((1,1,1,1,1)^{\top})$,
the projection of $\mathbf{q}_{2}$ onto the orthogonal complement of
$\hat{\alpha }_{1}$ is
$\mathbf{q}_{2}^{\perp}=(-0.63,-0.42,0.15,0.41,0.49)^{\top}$ (to 2
decimal places). Since $\Vert\mathbf{q}_{2}^{\perp}\Vert=0.999$, there
is no approximation to $\alpha_{2}$ orthogonal to $\hat{\alpha}_{1}$
which can
achieve an accuracy bigger than this. In particular, $\hat{\alpha}_{2}%
=\ell((1,1,0,-1,-1)^{\top})$ has an accuracy of $0.973$ with respect to
$\alpha_{2}$, while its accuracy with respect to
$\ell(\mathbf{q}_{2}^{\perp
})$ is slightly higher, being given by $\mathit{accu}(\alpha_{2},\hat{\alpha}%
_{2})/\Vert\mathbf{q}_{2}^{\perp}\Vert=0.973/0.999\approx0.974$.
Similar information for other axes is given in Table \ref{E2}.

Accordingly, to treat all axes of interest equally, we would in
principle consider all $k!$ possible orders. In practice, this can be
too many. Pragmatically, restricting attention to just the following
four orders has been found to work well. Together, they combine speed,
accuracy and a balance between prioritizing largeand small
eigenvalues, the two `next-best' orders incorporating an obvious greedy
heuristic:

\begin{description}
\item[Forwards (F):] Take the eigenaxes in decreasing order of their
eigenvalues.

\item[Backwards (B):] Take the eigenaxes in increasing order of their
eigenvalues.

\item[Next-best forwards (NF):] Take first the eigenaxis with the
largest eigenvalue and then, sequentially, the one with the largest
maximum possible achievable accuracy,
$\Vert\mathbf{q}_{r}^{\bot}\Vert$, among those remaining.

\item[Next-best backwards (NB):] Take first the eigenaxis with the
smallest eigenvalue and then, sequentially, the one with the largest
maximum possible achievable accuracy,
$\Vert\mathbf{q}_{r}^{\bot}\Vert$, among those remaining.
\end{description}
Whichever order is used to obtain them, the approximations
found are reported in the same $1$ to $k$ order.

To describe our approach in more detail, it suffices to consider a
single, fixed order. We use the forwards order below.

Note that when all eigenaxes are of interest ($k=p$), there is no
choice to be made when approximating the final axis, there being a
unique simple axis in $\mathbb{R}^{p}$ satisfying the $(p-1)$ orthogonality
requirements. In particular, the accuracy and complexity of the $p$th
axis approximation cannot be directly controlled. However, if the
first $(p-1)$ are accurate, then so too is the last.
Again, in general, the simpler the first $(p-1)$
approximations are, the simpler the last is. Overall, then, the number
of axes for which approximations are sought (rather than forced by
previous approximations) is $\widetilde{k}:=\min(k,p-1)$.

\subsection{Approximation for a given angle-accuracy}\label{sec:desc.app}

\subsubsection{Paradigm: Best $\theta$-accurate simple approximation}

We describe here the approximation paradigm at the heart of our
approach.

Recall that our approach favors simplicity over accuracy. Accordingly,
subject to being accurate enough---while orthogonal to previously
approximated axes---we seek the simplest possible approximation to
each axis in turn. If there is more than one such axis, we choose the
most accurate. More precisely, we adopt the paradigm: for a given angle
$\theta$, and for each $r=1,\ldots,\widetilde{k}$ in turn, seek the `best
$\theta$-accurate simple' approximation $\hat{\alpha}_{r}(\theta)$ to
$\alpha_{r}$ in the following sense.

For any $\theta\in(0,\pi/2)$, we say that an axis $\hat{\alpha}$ is
$\theta $-accurate for $\alpha_{r}$ if it is within an angle $\theta$\
of it---that is, if $\mathit{accu}(\hat{\alpha},\alpha_{r})>\cos(\theta)$. As
we have just seen, there are no such axes in $\mathcal{M}_{r}$ unless
$\cos(\theta)<\Vert \mathbf{q}_{r}^{\perp}\Vert$, so we always make
this requirement.

Again, we denote by $N_{r}(\theta)$ the smallest value of
$N\in\{1,2,\ldots \}$ for which there is a $\theta$-accurate axis in
$\mathcal{M}_{r}$ having complexity $N$. Thus, the `cone'
\[
C_{r}(\theta):=\{\hat{\alpha}\in\mathcal{M}_{r}\dvtx \mathit{accu}(\hat{\alpha}%
,\alpha_{r})>\cos(\theta) ,  \mathit{compl}(\hat{\alpha})=N_{r}(\theta)\}
\]
comprises all those axes in $\mathcal{M}_{r}$with the minimal
possible complexity $N_{r}(\theta)$ subject to being within an angle
$\theta$ of $\alpha_{r}$. For given $\theta$, we define `the best
$\theta$-accurate simple' approximation $\hat{\alpha}_{r}(\theta)$ to
$\alpha_{r}$ as the axis
in $C_{r}(\theta)$ closest to $\alpha_{r}$. That is, $\hat{\alpha}_{r}%
(\theta)$ \textit{is the closest of all the simplest possible,
}$\theta $ \textit{-accurate axes in} $\mathcal{M}_{r}$. Either of the
two possible integer representations of $\hat{\alpha}_{r}(\theta)$ will
be denoted by $\hat{\mathbf{z}}_{r}(\theta)$.

Finding $\hat{\alpha}_{r}(\theta)$ can be a hard combinatorial
optimization problem, especially when the dimension $p$ is large.
Therefore, to avoid the combinatorial complexity, we propose an
algorithm to approximate $\hat{\alpha }_{r}(\theta)$ which, after a
reordering of the variables, involves a computing effort linear in $p$,
for use when exact calculations are prohibitive. We briefly describe
such an algorithm in Appendix \ref{app:impl}.

We call $\cos(\theta)$ the minimum accuracy required for the
approximation to $\alpha_{r}$. We use the same value for each of the
$\widetilde{k}$ eigenaxes for which approximations are sought, denoting
by
$\hat{S}(\theta):=(\hat{\alpha}_{1}(\theta),\ldots,\hat{\alpha}_{k}(\theta))$
the full set of approximations obtained. To measure the overall
closeness of $\hat{S}(\theta)$ to $(\alpha_{1},\ldots,\alpha_{k})$, we
use the minimum of
the $k$ accuracies attained $ \{  \mathit{accu}(\alpha_{r},\hat{\alpha}_{r}%
(\theta)) \}  _{r=1}^{k}$, which we denote by
$\mathit{MA}(\hat{S}(\theta))$. As noted above, the larger this is, the smaller
a natural measure of overall distance between these two ordered sets of
axes (see, again, Appendix \ref{app:min.dist}).

\subsubsection{Tuning parameters}

Our approach uses the tuning parameters $N^{\ast}$ and~$\theta^{\ast}$,
described here, its results being typically less
sensitive to the choice of~$N^{\ast}$ due to its bias toward
simplicity. A third and final tuning parameter $\epsilon$, introduced
for operational convenience, is described in
Section \ref{sec:vary.theta}.

To facilitate interpretation, we require $N\leq N^{\ast}$, taking the
single digit default $N^{\ast}=9$ in all calculations reported here.
Thus, in practice, it may not be possible to complete the set of
approximations $\hat{S}(\theta)$, as some $N_{r}(\theta)$ may be found
to exceed~$N^{\ast}$. A similar effect occurs in \citet{Haus1982}
where, in effect, $N^{\ast}=1$.

As we want to stay close to the original eigenaxes, we require $\theta
\leq\theta^{\ast}$ for some $0<\theta^{\ast}\leq\pi/4$, values of
$\theta^{\ast}$ greater than $45^{\circ}$ clearly allowing poor
approximations. Thus, overall, we have the following bounds on the
accuracies attained for each $r=1,\ldots,\widetilde{k}$:
%
\begin{equation}\label{bounds accuracy}
\cos(\theta^{\ast})\leq\cos(\theta)<\mathit{accu}(\alpha_{r},\hat{\alpha}_{r}%
(\theta))\leq\Vert\mathbf{q}_{r}^{\perp}\Vert,
\end{equation}
where, for the first axis, we trivially have
$\mathbf{q}_{1}^{\perp }=\mathbf{q}_{1}$, so that
$\Vert\mathbf{q}_{1}^{\perp}\Vert=1 $. For most purposes, we recommend
taking $\theta^{\ast}=\pi/4$, an exhaustive account of all angles
smaller than $\theta^{\ast}$ being provided by considering an
automated sequence of angles, as described in
Section \ref{sec:vary.theta}. This choice of $\theta^{\ast}$ has the
advantage that no potentially useful approximations are ruled out of
consideration \textit{a priori}. Rather, the user is free to draw the
line regarding acceptable accuracy in the light of all the potentially
useful solutions found.

\subsubsection{Running example revisited}

We illustrate the above developments using our running example, with
$\theta=\theta^{\ast}=\pi/4$.

For the exams data, there is a $(\pi/4)$-accurate axis for $\alpha_{1}$
with complexity one; that is, $N_{1}(\pi/4)=1$. Further, out of all
axes of complexity one, $\ell((1,1,1,1,1)^{\top})$ is the closest to
$\alpha_{1}$. Therefore,
$\hat{\mathbf{z}}_{1}(\pi/4)=(1,1,1,1,1)^{\top}$ is an integer
representation of the best $(\pi/4)$-accurate simple approximation to~$\alpha_{1}$.

Here, $N_{2}(\pi/4)$ is also $1$, there being many $(\pi/4)$-accurate
axes for $\alpha_{2}$ with complexity one orthogonal to
$\hat{\alpha}_{1}(\pi/4)$, including $\ell((1,1,0,-1,-1)^{\top})$
and $\ell((1,0,0,0,-1)^{\top})$. Of these, we prefer the former, their
accuracies being 0.973 and 0.789, respectively. In fact, it can be
shown that $\ell((1,1,0,-1,-1)^{\top })$ is the best $(\pi/4)$-accurate
simple approximation to $\alpha_{2}$.

Again, $N_{3}(\pi/4)$ and $N_{4}(\pi/4)$ are also $1$, integer
representations of the corresponding best $(\pi/4)$-accurate simple
approximations being given in Table~\ref{E2} alongside $\hat{\mathbf{z}}%
_{1}(\pi/4)$ and $\hat{\mathbf{z}}_{2}(\pi/4)$. An extra decimal place
is used in reporting $\mathit{accu}(\alpha_{3},\hat{\alpha}_{3}(\pi/4))$ to show
where the minimum accuracy is attained.

As noted at the end of Section \ref{sec:seq.app}, there is no choice
about $\hat{\mathbf{z}}_{5}(\pi/4)$. However, illustrating the general
points made there, $\hat{\alpha}_{r}(\pi/4)$ being close to
$\alpha_{r}$ for each $r=1,\ldots,4$, $\hat{\alpha}_{5}(\pi/4)$ is also
close to $\alpha_{5}$ (having an accuracy of $0.974$), while the
relative simplicity of $\hat{\alpha}_{5}$ reflects that of
$\hat{\alpha}_{1}$ to $\hat{\alpha}_{4}$.

\subsection{Effect of varying the minimum accuracy required}\label{sec:vary.theta}

When $\theta=\pi/4$ the approximations $\hat{S}(\theta)$ typically have
low complexity overall and so can usually be interpreted. Unless all
the eigenaxes are already simple, we might expect the overall
complexity of the approximations to steadily increase with the minimum
accuracy required. However, it turns out there is no straightforward
relationship between the complexity of the approximations and $\theta$.
This nonmonotone behavior of the approximations $\hat{S}(\theta)$ when
$\cos(\theta)$ increases is due to the discreteness inherent in our
approximations. Restricting the elements of the integer representations
to be coprime is mainly responsible for this, division by a highest
common factor greater than $1$ always being a possibility.

The net effect is that it is not possible to fully predict the
qualitative behavior of $\hat{S}(\theta)$ as $\theta$ varies.
Accordingly, instead of attempting to find an optimal value of
$\cos(\theta)$ under some criterion, we vary the value of $\theta$ so
as to explore \textit{all} possible sets of approximations. The
different sets of orthogonal axes thereby obtained offer different
views of the same data set, giving the user more scope for
interpretation.

The good news is that it is only necessary to explore a discrete set of
values of~$\theta$. To see this, we introduce the following notation.
For any $0<\theta\leq\theta^{\ast}$, we denote by
\[
\widetilde{S}(\theta):=\bigl(\hat{\alpha}_{1}(\theta),\ldots,\hat{\alpha
}_{\widetilde{k}(\theta)}(\theta)\bigr),\qquad\mbox{where }\widetilde{k}(\theta
)\leq\widetilde{k},%
\]
the ordered set of approximate axes obtained among the $\widetilde{k}%
=\min(k,p-1)$ sought. This set is complete [$\widetilde{k}(\theta
)=\widetilde{k} $] unless there is a first $\widetilde{k}(\theta
)<\widetilde{k}$ with $N_{\widetilde{k}(\theta)+1}(\theta)$ found to
be greater than $N^{\ast}$. When $\widetilde{S}(\theta)$ is complete,
so too is the full set of $k$ approximate axes $\widehat{S}(\theta)$,
being given by
\[
\widehat{S}(\theta)= \cases{
\widetilde{S}(\theta), &\quad if $k<p$,\cr
(\widetilde{S}(\theta),\hat{\alpha}_{p}(\theta)), &\quad if $k=p$,%
}
\]
where $\hat{\alpha}_{p}(\theta)$ is the unique simple axis in $\mathbb{R}%
^{p}$ orthogonal to the $(p-1)$ axes in $\widetilde{S}(\theta)$.
Otherwise, $\widehat{S}(\theta)$ itself is incomplete, and so not
reported. In all cases, the minimum accuracy attained among the axes in
$\widetilde{S}(\theta)$, denoted $\mathit{MA}(\widetilde{S}(\theta))$, satisfies
$\mathit{MA}(\widetilde{S}(\theta ))>\cos(\theta)$, by (\ref{bounds accuracy}).
For any $0<\theta\leq \theta^{\ast}$, defining $\theta^{+}<\theta$ by
\[
\cos(\theta^{+})=\mathit{MA}(\widetilde{S}(\theta)),%
\]
it follows that the same set of approximations is obtained [$\widetilde
{S}(\theta)=\widetilde{S}(\theta^{\prime})$] for all smaller angles
$\theta^{\prime}$ in the range $(\theta^{+},\theta)$ determined by
\[
\cos(\theta)<\cos(\theta^{\prime})<\cos(\theta^{+}),%
\]
but that change happens at the more accurate end of this range, $\theta^{+}%
$-accuracy precluding
$\widetilde{S}(\theta)=\widetilde{S}(\theta^{+})$.

Thus, to \textit{fully} explore the range of possible approximations,
it is sufficient to consider the strictly decreasing sequence of angles
$\theta^{[1]},\theta^{[2]},\ldots$ defined by%
%
\begin{equation}\label{sequence angles full}
\theta^{[1]}:=\theta^{\ast}\quad\mbox{and}\quad\theta^{[n+1]}%
:=(\theta^{[n]})^{+}\qquad(n\geq1).%
\end{equation}
In practice, for operational convenience, we stop when the minimum
accuracy required $\cos(\theta)$ reaches $(1-\epsilon)$ for some small
tuning parameter $\epsilon$. In general, no simple solutions are missed
by doing this, approximations with very high minimum accuracy required
usually being very complex. Experience has shown that a value of
$\epsilon=0.01$ gives satisfactory results, while also keeping the
computations fast.

Key features of the relation between consecutive sets of approximations
obtained, $\widetilde{S}(\theta^{[n]})$ and $\widetilde{S}%
(\theta^{[n+1]})$, now follow. Let
%
\begin{equation}
r_{n}:=\arg  \min_{1\leq r\leq\widetilde{k}(\theta^{[n]})}%
\mathit{accu}\bigl(\alpha_{r},\hat{\alpha}_{r}\bigl(\theta^{[n]}\bigr)\bigr)
\end{equation}
indicate the first approximation which changes from $n$ to $n+1$.
Earlier approximated axes do not change as
$\{\hat{\alpha}_{r}(\theta^{[n]})\}_{r=1}^{r_{n}-1}$ are already
$\theta^{[n+1]}$-accurate. However, for $\alpha_{r_{n}}$ an
approximation strictly more accurate than $\hat
{\alpha}_{r_{n}}(\theta^{[n]})$ must be sought. Further, if
$k=p$ while $\hat{S}(\theta^{[n]})$ and
$\hat{S}(\theta^{[n+1]})$ are complete, the orthogonality
restrictions imply that the subspace generated by the remaining
approximate axes is the same for $\hat{S}(\theta^{[
n+1]})$ as it is for $\hat{S}(\theta^{[n]})$. That is,%
\[
\operatorname{span}\bigl\{\hat{\mathbf{z}}_{r}\bigl(\theta^{[n+1]}\bigr)\dvtx r=r_{n}%
,\ldots,p\bigr\}=\operatorname{span}\bigl\{\hat{\mathbf{z}}_{r}\bigl(\theta^{[n]}%
\bigr)\dvtx r=r_{n},\ldots,p\bigr\}.%
\]
In other words, we are obtaining a different, more accurate, orthogonal
simple basis for the same subspace.

\begin{table}
\caption{Integer representations for the examinations data with
$\cos(\theta^{[3]})=0.9375$}\label{E22}%
\begin{tabular*}{\textwidth}{@{\extracolsep{\fill}}lccccc@{}}
\hline
\textbf{Variable} & $\bolds{\hat{\mathbf{z}}_{1}(\theta)}$ &
$\bolds{\hat{\mathbf{z}}_{2}(\theta)}$ & $\bolds{\hat{\mathbf{z}}_{3}(\theta)}$ &
$\bolds{\hat{\mathbf{z}}_{4}(\theta)}$ & $\bolds{\hat{\mathbf{z}}_{5}(\theta)}$\\
\hline
Mechanics (closed) & 1\phantom{000} & 1\phantom{000} & 2\phantom{000} & 1\phantom{000} & 1\phantom{0}\\
Vectors (closed) & 1\phantom{000} & 1\phantom{000} & $-$2\phantom{0000,} & $-$1\phantom{0000,} & 1\phantom{0}\\
Algebra (open) & 1\phantom{000} & 0\phantom{000} & 0\phantom{000} & 0\phantom{000} & $-$4\phantom{00,}\\
Analysis (open) & 1\phantom{000} & $-$1\phantom{0000,} & $-$1\phantom{0000,} & 2\phantom{000} & 1\phantom{0}\\
Statistics (open) & 1\phantom{000} & $-$1\phantom{0000,} & 1\phantom{000} & $-$2\phantom{0000,} & 1\phantom{0}\\ [6pt]
Accuracy & 0.997 & \textbf{0.973} & 0.980 & 0.979 & \phantom{00}0.974\\ [3pt]
Variance (\%)& 63.3\phantom{000} & 14.4\phantom{000} & 8.9\phantom{00} & 7.8\phantom{00} & 5.5\\
\hline
\end{tabular*}
\end{table}

With $\theta^{[1]}=\theta^{\ast}=\pi/4$, the minimum accuracy
required is $1/\sqrt{2}\approx0.7071$. For the exams data, the minimum
accuracy \textit{attained} in this case is for the fourth eigenaxis, so
that $r_{1}=4$ and $\cos(\theta^{[2]})=0.937$ (see
Table \ref{E2}). We therefore have
at once that $\hat{\alpha}_{r}(\theta^{[2]})=\hat{\alpha}_{r}%
(\theta^{[1]})$ for $r=1,2$ and $3$. However, it is not possible
to find an improved accuracy approximation
$\hat{\alpha}_{4}(\theta^{[2]})$ with
complexity at most $N^{\ast}=9$, so that $\widetilde{k}(\theta^{[2]}%
)=3$ and $\widetilde{S}(\theta^{[2]})$ is incomplete. Without
further calculation, Table \ref{E2} gives $r_{2}=3$,
$\cos(\theta^{[
3]})=0.9375$ and $\hat{\alpha}_{r}(\theta^{[3]})=\hat{\alpha}_{r}%
(\theta^{[2]})=\hat{\alpha}_{r}(\theta^{[1]})$ for $r=1$
and~$2$.

In fact, $\hat{S}(\theta^{[3]})$ is complete, corresponding
integer representations being reported in Table \ref{E22}. The
increase in minimum accuracy required in going from
$\theta^{[2]}$ to $\theta^{[3]}$ is very small but, due to
discreteness effects, results in a \textit{drop} in the complexities of
the third and fourth axis approximations down below $N^{\ast}=9$. The
newly approximated axes $\{\hat{\alpha}_{r}(\theta
^{[3]})\}_{r=3}^{5}$ span the same subspace as $\{\hat{\alpha}%
_{r}(\theta^{[1]})\}_{r=3}^{5}$. Further, $\hat{\alpha}_{5}%
(\theta^{[3]})=\hat{\alpha}_{5}(\theta^{[1]})$ precisely
because,
in this example, $\{\hat{\alpha}_{3}(\theta^{[3]}),\hat{\alpha}%
_{4}(\theta^{[3]})\}$ and $\{\hat{\alpha}_{3}(\theta^{[1]}%
),\hat{\alpha}_{4}(\theta^{[1]})\}$ span the same two-dimensional
subspace. The $\hat{S}(\theta^{[3]})$ pair of axes here are now
almost as simple as those for $\hat{S}(\theta^{[1]})$ but more
accurate, striking a different simplicity-accuracy trade-off. Comparing
Tables \ref{E2} and \ref{E22}, this example also illustrates that
moving to a new simplicity-accuracy trade-off need not change the
variances explained by each axis in any material way.

From $\cos(\theta^{[4]})=0.973$ onward, it is not possible to
find complete sets of approximations $\hat{S}(\theta)$ with complexity
at most $N^{\ast}=9$. Thus, for the forwards order of approximation,
Tables \ref{E2} and \ref{E22}, detailing
$\hat{S}(\theta^{[1]})$ and $\hat {S}(\theta^{[3]})$,
respectively, together cover the full range
$0<\theta\leq\theta^{\ast}=\pi/4$.

\subsection{Automated visual display of solutions}\label{sec:auto}

Based on $\mathcal{S}$, the complete\vspace*{-1pt} set of solutions (sets of
approximate axes) $\hat{S}(\theta)$ found\vspace*{2pt} by one or more of the four
orders of approximation described in Section \ref{sec:seq.app}, the
user can now proceed\vspace*{2pt} to answer the open question posed at the outset:
\begin{quote}
What sets of simply interpretable orthogonal axes---if
any---are  angle-close to the principal components of
interest?
\end{quote}
In principle, an overall informed choice requires the user to
compare all solutions regarded as angle-close in terms of a range of
factors, including subject matter considerations, as described in
Section \ref{sec:inter}. Only the individual user can calibrate the
various trade-offs involved and different users will, quite reasonably,
choose different (numbers of) solutions.

In practice, the work involved can be substantial and, to help the user
make this choice, we provide a summary automated visual display in
which the solutions found are ordered in terms of overall measures of
star quality, simplicity and accuracy. Tabular information for each
solution is presented to the user in this order. In describing this
automated display here, we emphasize that---although clearly
principled---this order of solutions does not, indeed cannot, presume
to be the preference order for any particular user.

\subsubsection{Accuracy--simplicity scatterplot}\label{sec:accsimp.plot}

Prioritizing our main criteria of simplicity and accuracy, each
solution $\hat{S}$ is plotted at a point in the positive quadrant with
the following
coordinates. Horizontally, we use the discrepancy measure
\[
\mathit{discr}(\hat{S})=1-\mathit{MA}(\hat{S}),
\]
a natural measure of (squared) distance between the eigenaxes
$\mathbf{(\pm q}_{1}|\cdots|\mathbf{\pm q}_{k}\mathbf{)}$ and~$\hat{S}$.
The smaller $\mathit{discr}(\hat{S})$, the more accurate $\hat{S}$. Vertically,
we use overall
complexity measure%
\[
\mathit{compl}(\hat{S})=N_{\max}(\hat{S})+\lambda(\hat{S}),
\]
where $N_{\max}(\hat{S})$ is the maximum complexity of the axes in
$\hat{S}$, the term
$0<\lambda(\hat{S}):=\frac{\sqrt{\sum _{h}\sum_{r}\widehat
{z}_{hr}^{2}/(pk)}}{2N_{\max}(\hat{S})}\leq\frac{1}{2}$ being included
to further discriminate between solutions with the same maximum
complexity. The smaller $\mathit{compl}(\hat{S})$, the simpler $\hat{S}$.

\begin{figure}

\includegraphics{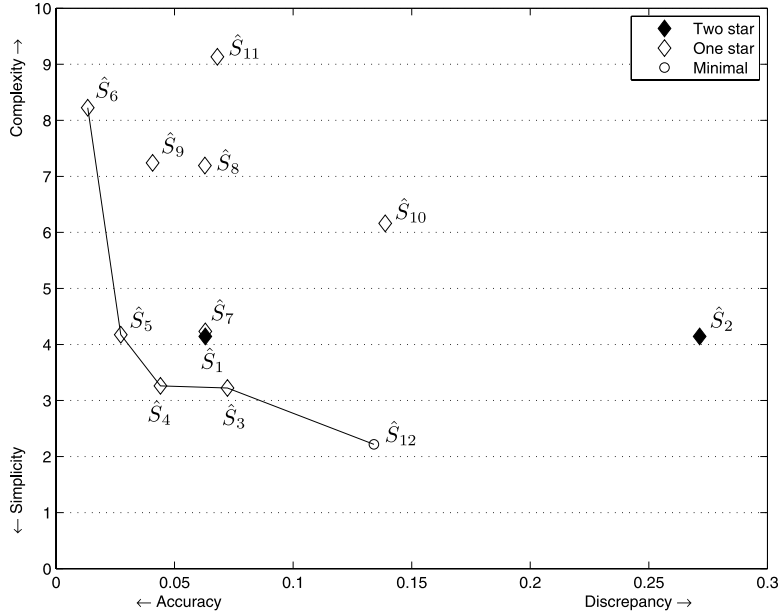}

\caption{Solution set $\mathcal{S}$ for the exams data.}\label{fig:exams}
\end{figure}

Figure \ref{fig:exams} shows the corresponding scatterplot for the
exams data where, overall, 12 different solutions were obtained. The
numbering of the solutions reflects a particular principled order
described below, together with the plot symbols used. In particular,
the forwards solutions $\hat {S}(\theta^{[1]})$ and
$\hat{S}(\theta^{[3]})$ discussed above appear here as
$\hat{S}_{1}$ and $\hat{S}_{7}$, respectively.

\subsubsection{Minimal and dominated solutions}

Low values of both $\mathit{discr}(\hat{S})$ and $\mathit{compl}(\hat{S})$ are clearly
desirable, but cannot usually be simultaneously achieved. For example,
Figure \ref{fig:exams} shows that there is a trade-off for the exams
data, with no solution attaining the smallest value of both these
coordinates. However, the five solutions joined by straight lines are
visibly special, the rectangle formed by each of them with the origin
containing no other solutions. For any set of solutions, we call these
the \textit{minimal} solutions---those for which no lower value of
either coordinate can be found without increasing the other. Thus,
here, among $\hat{S}_{3},\hat{S}_{4},\hat{S}_{5},\hat{S}_{6}$ and
$\hat{S}_{12}$, the simpler solutions are less accurate, and the more
accurate solutions are less simple.

There is always at least one minimal solution, usually more. Together,
they form the lower-left boundary of the scatterplot whose shape
reflects, in any particular case, the trade-off between simplicity and
accuracy. All other\vspace*{1pt} solutions are \textit{dominated} by a minimal
solution. For example, here,
$\hat{S}_{1}$ and $\hat{S}_{7}$ are dominated by $\hat{S}_{4}$, with $\hat{S}_{4}%
$ being simpler and more accurate than both.

\subsubsection{Star quality solutions}

In general, focusing only on solutions which lie in the minimal set
does not necessarily capture all clearly interpretable solutions.
Other, dominated, solutions may possess `star quality' in the sense
that there are striking overall patterns in the set of approximate axes
found, deserving to be especially drawn to the user's attention. For
example, $\hat{S}_{1}$ (Table \ref{E2}, repeated here as the left-hand
part of Table \ref{two stars solutions books data}) has a very clear
and interpretable structure and so deserves to be brought early to the
user's attention, even though it does not lie in the minimal set. For
many users this increased interpretability is likely to be worth the
cost in terms of discrepancy and overall complexity.

We recognize that interpretability is a subjective concept. Rather than
attempting a quantification, we use a star rating system to indicate
the degree to which a solution conforms with one of a predefined set of
clear structures: `two star' solutions conform to the clearest
structures and `one star' to the next clearest, while `unstarred'
solutions do not conform to any of the predefined structures.

Here, we use six predefined structures, these being one and two star
versions of three mutually exclusive types, denoted \textbf{A},
\textbf{B} and \textbf{C}, described next. There are clear points of
contact with the work of Rousson and Gasser, summarized in the
following Section \ref{sec:rg}.

\begin{table}\tablewidth=330pt
\tabcolsep=0pt
\caption{Integer representations of the two star solutions for the
exams data}\label{two stars solutions books data}
\begin{tabular*}{330pt}{@{\extracolsep{\fill}}lk{2.3}k{2.3}k{1.4}k{1.3}k{1.3}k{2.3}k{2.3}k{1.4}k{1.3}k{1.3}@{}}
\hline
&\multicolumn{5}{c}{$\bolds{\hat{\bolds{S}}_{1}}$} \hspace*{6pt}& \multicolumn{5}{c@{}}{$\bolds{\hat{\bolds{S}}_{2}}$}\\ [-7pt]
&\multicolumn{5}{c}{\hrulefill}&\multicolumn{5}{c@{}}{\hrulefill}\\
\multicolumn{1}{@{}l}{\textbf{Variable}} & \multicolumn{1}{c}{$\bolds{\hat{\mathbf{z}}_{1}}$\phantom{,}} & \multicolumn{1}{c}{$\bolds{\hat{\mathbf{z}}_{2}}$\phantom{0}}
& \multicolumn{1}{c}{$\bolds{\hat{\mathbf{z}}_{3}}$\phantom{0}}
& \multicolumn{1}{c}{$\bolds{\hat{\mathbf{z}}_{4}}$\phantom{0}} & \multicolumn{1}{c}{$\bolds{\hat{\mathbf{z}}_{5}}$} \hspace*{6pt}&
\multicolumn{1}{c}{$\bolds{\hat{\mathbf{z}}_{1}}$}
& \multicolumn{1}{c}{$\bolds{\hat{\mathbf{z}}_{2}}$\phantom{0}} & \multicolumn{1}{c}{$\bolds{\hat{\mathbf{z}}_{3}}$\phantom{0}}
& \multicolumn{1}{c}{$\bolds{\hat{\mathbf{z}}_{4}}$\phantom{0}} & \multicolumn{1}{c@{}}{$\bolds{\hat{\mathbf{z}}_{5}}$\phantom{0}}\\
\hline
Mechanics & 1. & 1. & 1. & 0. & 1. & 1. & 1. & 1. & 1. & 0.\\
Vectors & 1. & 1. & -1. & 0. & 1. & 1. & 1. & -1. & 1. & 0.\\
Algebra & 1. & 0. & 0. & 0. & -4. & 1. & -1. & 0. & 1. & 1.\\
Analysis & 1. & -1. & 0. & 1. & 1. & 1. & -1. & 0. & 1. & -1.\\
Statistics & 1. & -1. & 0. & -1. & 1. & 1 & 0. & 0. & -4. & 0.\\ [6pt]
Accuracy & \multicolumn{1}{d{2.3}}{0.997} & \multicolumn{1}{d{2.3}}{0.973} & \multicolumn{1}{d{1.4}}{0.9375}
& \multicolumn{1}{d{1.3}}{0.937} & \multicolumn{1}{d{1.3}}{0.974} & \multicolumn{1}{d{2.3}}{0.997}
& \multicolumn{1}{d{2.3}}{0.802} & \multicolumn{1}{d{1.4}}{0.9375} &\multicolumn{1}{d{1.3}}{0.729}
& \multicolumn{1}{d{1.3}@{}}{0.897}\\ [3pt]
 Variance (\%)& \multicolumn{1}{d{2.3}}{63.3} & \multicolumn{1}{d{2.3}}{14.4} & \multicolumn{1}{d{1.4}}{8.9}
& \multicolumn{1}{d{1.3}}{7.9} & \multicolumn{1}{d{1.3}}{5.5}& \multicolumn{1}{d{2.3}}{63.3}
& \multicolumn{1}{d{2.3}}{12.1} & \multicolumn{1}{d{1.4}}{8.9} & \multicolumn{1}{d{1.3}}{9.9} & \multicolumn{1}{d{1.3}}{5.8}\\\hline
\end{tabular*}
\end{table}

Let $\widehat{\mathbf{Z}}=(\hat{\mathbf{z}}_{1}|\cdots|\hat{\mathbf{z}%
}_{k})$ be a matrix of integer representations of $\hat{S}$. Each
structure used requires that the $p$ variables can be partitioned into
$b\geq1$ blocks, where each block labels the set of nonzero (by
convention, positive) elements of a single-signed column
$\hat{\mathbf{z}}_{r}$. If $b<k$, orthogonality entails that the
remaining $(k-b)$ columns of $\widehat{\mathbf{Z}}$ are contrasts---that
is, have elements of both signs. The within-block condition
\textbf{W-B} [condition 3 in \citet{RousGass2004}] holds if the
nonzero elements of each contrast occur within a single block.

In this terminology, the three mutually exclusive types of predefined
structure are as follows:
\begin{description}
\item[A:] $b=1$---that is, an overall (possibly, weighted)
mean, plus orthogonal contrasts.

\item[B:] $b>1$ and \textbf{W-B }holds, so that each block has type
\textbf{A} structure.

\item[C:] $b>1$ and \textbf{W-B }does not hold.
\end{description}
The type of each starred solution is noted in its table of
information but, for visual clarity, not in the accuracy--simplicity
scatterplot.

We call $\hat{\mathbf{z}}_{r}$ \textit{parsimonious} if
$N_{r}^{\sharp}$, the number of distinct nonzero elements it contains,
is small. The more parsimonious a starred solution, the clearer its
structure. Accordingly, we award two stars when it is as parsimonious
as possible of its type, and one star otherwise. Orthogonality entails
that, for each type, two star solutions
are precisely those which obey the following two conditions:
\begin{eqnarray*}
\max\{N_{r}^{\sharp}\dvtx\hat{\mathbf{z}}_{r}\mbox{ defines a block}%
\}&=&1\quad\mbox{and}\\
\max\{N_{r}^{\sharp}\dvtx\hat{\mathbf{z}}_{r}\mbox{ defines a
contrast}\}&=&2.
\end{eqnarray*}
For example, adopting an obvious notation, an \textbf{A}$^{\ast\ast}%
$ solution has a simple arithmetic mean, plus a set of orthogonal
contrasts in each of which the nonzero elements comprise $m$ times a
value $n$, and $n$ times a value $-m$, whereas an \textbf{A}$^{\ast}$\
solution has either an unequally weighted mean, or a contrast not of
this form.

Examples of the six possible starred structures are given below, the
\textbf{A}$^{\ast}$ example being $\hat{S}_{3}$ in
Figure \ref{fig:exams} and detailed in the left-hand part of Table
\ref{one stars solutions books data}.\\
\begin{center}
\tabcolsep=0pt
\begin{tabular*}{\textwidth}{@{\extracolsep{\fill}}lccc@{}}
\hline
& \multicolumn{3}{c@{}}{\textbf{Structure type}}\\ [-7pt]
&\multicolumn{3}{c@{}}{\hrulefill}\\
&\textbf{A} & \textbf{B} & \textbf{C}\\
\hline
Two star &
$ {\fontsize{8.36}{10}\selectfont{\left(\matrix{
1 & \phantom{-}1 & \phantom{-}1 &\phantom{-}0 & \phantom{-}1\cr
1 & \phantom{-}1 & -1 & \phantom{-}0 & \phantom{-}1\cr
1 & \phantom{-}0 & \phantom{-}0 & \phantom{-}0 & -4\cr
1 & -1 & \phantom{-}0 & \phantom{-}1 & \phantom{-}1\cr
1 & -1 & \phantom{-}0 & -1 & \phantom{-}1
}\right)}}$&
$ {\fontsize{8.36}{10}\selectfont{\left(\matrix{
1 & 0 & \phantom{-}1 & \phantom{-}0\cr
1 & 0 & -1 & \phantom{-}0\cr
0 & 1 & \phantom{-}0 & \phantom{-}1\cr
0 & 1 & \phantom{-}0 & -1
}\right)}}$ &
$ {\fontsize{8.36}{10}\selectfont{\left(\matrix{
1 & 0 & \phantom{-}1 & \phantom{-}1\cr
1 & 0 & -1 & -1\cr
0 & 1 & \phantom{-}1 & -1\cr
0 & 1 & -1 & \phantom{-}1
}\right)}} $\\
\hline One star &
$ {\fontsize{8.36}{10}\selectfont{\left(\matrix{
3 & -1 & \phantom{-}1 & \phantom{-}0 & \phantom{-}0\cr
3 & -1 & -1 & \phantom{-}0 & \phantom{-}0\cr
2 & \phantom{-}1 & \phantom{-}0 & \phantom{-}0 & -2\cr
2 & \phantom{-}1 & \phantom{-}0 & \phantom{-}1 & \phantom{-}1\cr
2 & \phantom{-}1 & \phantom{-}0 & -1 & \phantom{-}1
}\right)}}$ &
${\fontsize{8.36}{10}\selectfont{ \left(\matrix{
1 & 0 & \phantom{-}2 & \phantom{-}0\cr
2 & 0 & -1 & \phantom{-}0\cr
0 & 1 & \phantom{-}0 & \phantom{-}2\cr
0 & 2 & \phantom{-}0 & -1
}\right)}}$ &
$ {\fontsize{8.36}{10}\selectfont{\left(\matrix{
1 & 0 & \phantom{-}2 & \phantom{-}2\cr
2 & 0 & -1 & -1\cr
0 & 1 & \phantom{-}2 & -2\cr
0 & 2 & -1 &\phantom{-}1
}\right)}}$ \\
\hline
\end{tabular*}
\end{center}

\subsubsection{Empirical support for assumed models: Points of contact with
Rousson and Gasser}\label{sec:rg}

Our approach seeks solutions supported by the data in the sense that no
modeling assumptions are imposed, apart from our axes being orthogonal
and containing vectors of integers. Therefore, if an optimal solution
under some modeling assumptions is produced by our analysis, this
provides empirical evidence in favor of such a model.

We develop this general point here with respect to the Rousson and
Gasser method, as reported in \citet{RousGass2004}, with which there
are clear points of contact, recalling that it applies to correlation
matrices only.

\begin{table}\tablewidth=330pt
\tabcolsep=0pt
\caption{Integer representations of the two best one star
approximations for exams data}\label{one stars solutions books data}
\begin{tabular*}{330pt}{@{\extracolsep{\fill}}lk{2.3}k{2.3}k{1.3}k{1.3}k{1.3}k{2.3}k{2.3}k{1.2}k{1.3}k{1.3}@{}}
\hline
&\multicolumn{5}{c}{$\bolds{\hat{S}_{1}}$} \hspace*{6pt}& \multicolumn{5}{c@{}}{$\bolds{\hat{S}_{2}}$}\\ [-7pt]
&\multicolumn{5}{c}{\hrulefill}&\multicolumn{5}{c@{}}{\hrulefill}\\
\multicolumn{1}{@{}l}{\textbf{Variable}} & \multicolumn{1}{c}{$\bolds{\hat{\mathbf{z}}_{1}}$\phantom{,}} &
\multicolumn{1}{c}{$\bolds{\hat{\mathbf{z}}_{2}}$\phantom{0}}
& \multicolumn{1}{c}{$\bolds{\hat{\mathbf{z}}_{3}}$}
& \multicolumn{1}{c}{$\bolds{\hat{\mathbf{z}}_{4}}$} & \multicolumn{1}{c}{$\bolds{\hat{\mathbf{z}}_{5}}$} \hspace*{6pt}&
\multicolumn{1}{c}{$\bolds{\hat{\mathbf{z}}_{1}}$}
& \multicolumn{1}{c}{$\bolds{\hat{\mathbf{z}}_{2}}$\phantom{0}} & \multicolumn{1}{c}{$\bolds{\hat{\mathbf{z}}_{3}}$\phantom{0}}
& \multicolumn{1}{c}{$\bolds{\hat{\mathbf{z}}_{4}}$\phantom{0}} & \multicolumn{1}{c@{}}{$\bolds{\hat{\mathbf{z}}_{5}}$\phantom{0}}\\
\hline
Mechanics & 3. & -1. & 1. & 0. & 0. & 1. & 3. & 2. & 1. & 0.\\
Vectors & 3. & -1. & -1. & 0. & 0. & 1. & 3. & -2. & -1. & 0.\\
Algebra & 2. & 1. & 0. & 0. & -2. & 1. & -2. & 0. & 0. & -2.\\
Analysis & 2. & 1. & 0. & 1. & 1. & 1. & -2. & -1. & 2. & 1.\\
Statistics & 2. & 1. & 0. & -1. & 1. & 1 & -2. & 1. & -2. & 1.\\ [6pt]
Accuracy & \multicolumn{1}{d{2.3}}{0.996} & \multicolumn{1}{d{2.3}}{0.928} & \multicolumn{1}{d{1.3}}{0.937}
& \multicolumn{1}{d{1.3}}{0.937} & \multicolumn{1}{d{1.3}}{0.959} & \multicolumn{1}{d{2.3}}{0.997}
& \multicolumn{1}{d{2.3}}{0.956} & \multicolumn{1}{d{1.2}}{0.98} &\multicolumn{1}{d{1.3}}{0.978}
& \multicolumn{1}{d{1.3}@{}}{0.959}\\ [3pt]
Variance (\%)& \multicolumn{1}{d{2.3}}{60.2} & \multicolumn{1}{d{2.3}}{17.3} & \multicolumn{1}{d{1.3}}{8.9}
& \multicolumn{1}{d{1.3}}{7.9} & \multicolumn{1}{d{1.3}}{5.7}& \multicolumn{1}{d{2.3}}{63.3}
& \multicolumn{1}{d{2.3}}{14.2} & \multicolumn{1}{d{1.2}}{8.9} & \multicolumn{1}{d{1.3}}{7.8} & \multicolumn{1}{d{1.3}}{5.7}\\\hline
\end{tabular*}
\end{table}

A key point is that two star structures emerging from our essentially
exploratory analysis of the data correspond to assumed structures
optimally fitted to it in \citet{RousGass2004}, with the added
condition of orthogonality between \textit{all} pairs of approximate
axes. Accordingly, solutions generated by \citet{RousGass2004} can
only coincide with ours when they are orthogonal, in which case they
are two star solutions. In particular, one star solutions---involving
weighted means and/or contrasts not of the \textbf{A}$^{\ast\ast}$\
form noted above---cannot arise in such an analysis.

For any given number of blocks $b$, the scalar target function
optimized in \citet{RousGass2004}---the `corrected sum of variances'\
[see the paper by \citet{GervRous2004}], denoted here by
$\mathit{optim}(\hat{S})$---can be used whether components are orthogonal or
not. Although not optimized for in our approach, its value can be
calculated for each of our solutions and compared to the best value
found in \citet{RousGass2004}. However, its maximization reflects an
exclusive interest in explaining variability, whereas, being as
interested in exploring potential scientific laws, our approach treats
all eigenvectors of interest equally.

Overall, the two methods will give complementary results, agreement
only being expected when there is strong empirical evidence of an
orthogonal two star structure underpinning the variability in the data.

\subsubsection{A total order of solutions}

A principled total order of the full set of solutions $\mathcal{S}$ is
now obtained, in two stages.

First, we place each solution into one of four classes, ordered by
interpretability: two star solutions, one star solutions, unstarred
solutions lying in the minimal set for $\mathcal{S}$, and the rest. All
solutions in a higher class are ranked ahead of all those in a lower
class.

Then, we order the solutions within each class by simplicity and
accuracy, with our usual bias toward the former, as follows. Having
found the minimal set \textit{for a~given class}, we give its solutions
the highest available rankings, ordering them by $\mathit{compl}(\hat{S})$ [any
ties being broken by $\mathit{discr}(\hat{S})$]---in other words, working from
right to left in the accuracy--simplicity scatterplot. We now remove
this minimal set from the class and repeat the ranking procedure on the
remainder until all solutions in the class have been ranked.

Tabular information for each solution is presented to the user in the
resulting total order. For visual clarity, solutions in the lowest
class---those neither starred, nor minimal for $\mathcal{S}$---are
not numbered in the scatterplot.

For example, for the exams data, the two star class comprises
$\hat{S}_{1}$ and $\hat{S}_{2}$ (shown in Table \ref{two stars
solutions books data}), reflecting the fact that they are considered
top in terms of interpretability. Both are of type \textbf{A}. Between
them, $\hat{S}_{1}$ is ranked higher as it dominates $\hat{S}_{2}$,
having the same overall complexity but a much better minimum accuracy
attained: $0.937$ compared to $0.729$ (for the fourth axis in both
cases). Indeed, the corresponding angle for this axis being some
$43^{\circ}$, it seems likely that many users will rule out
$\hat{S}_{2}$ as being insufficiently accurate.

The one star class comprises solutions $\hat{S}_{3}$ to $\hat{S}_{11}$.
Among them, solutions $\hat{S}_{3}$--$\hat{S}_{6}$ have the highest
rankings since they form the corresponding minimal set---that
is, within this class, it is not possible to improve on either overall
simplicity or accuracy without doing worse on the other criterion. They
are ranked by overall simplicity [small values of $\mathit{compl}(\hat{S})$].
Removing them from the class and continuing, the new (ordered) minimal
classes are $\hat{S}_{7}$ to $\hat {S}_{9}$ and, finally,
$\hat{S}_{10}$ and $\hat{S}_{11}$. For the exams data, $\hat{S}_{12}$
is the only unstarred solution in the minimal set for $\mathcal{S}$,
while there are no solutions in the lowest class.

\subsection{Running example: Summary comparison of results}\label{sec:exams.res}

We summarize here our results for the exam data running example
(Section \ref{sec:org}) displayed in Figure~\ref{fig:exams}, whose
terminology is explained above, comparing them with those of
\citet{RousGass2004} and \citet{Vine2000}.


Overall, the user is referred first to $\hat{S}_{1}$, shown in the left
part of Table~\ref{two stars solutions books data}. This effectively
combines simplicity, accuracy and subject matter interpretability, this
latter being particularly straightforward:

\begin{longlist}
\item[$\hat{\alpha}_{1}$:] Represents overall mathematical ability.

\item[$\hat{\alpha}_{2}$:] Contrasts closed- and open-book exam
performance, omitting \mbox{Algebra.}

\item[$\hat{\alpha}_{3}$:] Contrasts performance in the two closed-book
exams, Mechanics and Vectors.

\item[$\hat{\alpha}_{4}$:] Contrasts performance in the two open-book
exams, Analysis and Statistics, included in $\hat{\alpha}_{2}$.

\item[$\hat{\alpha}_{5}$:] Contrasts Algebra with all other subjects.
\end{longlist}

\begin{figure}
\vspace*{20pt}

\includegraphics{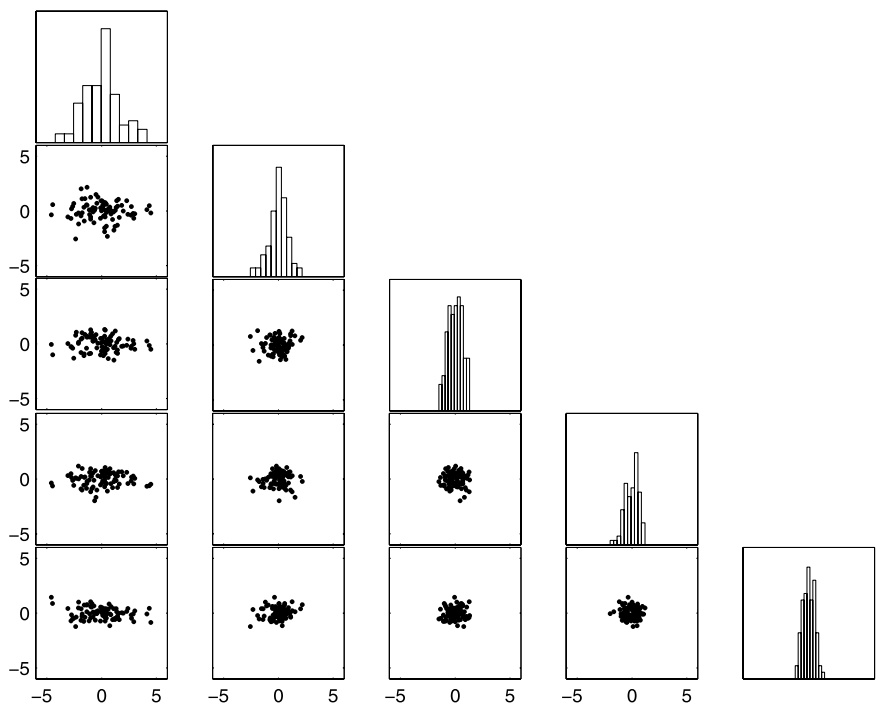}

\caption{Scatterplot matrix of the simplified principal components for
the exams data.} \label{fig:exams.scatter}
\end{figure}

Figure \ref{fig:exams.scatter} shows the scatterplot matrix for $\hat
{S}_{1}$, the visualization and dimension reduction features offered by
such orthogonal-axis plots only being guaranteed with rotation
approaches such as ours. The performance of each student on each of
these five readily interpreted axes is visible. In particular, two or
three students stand out at either extreme of overall mathematical
ability, these students having very similar open- and closed-book
performances as measured by $\hat{\alpha }_{2}$. Again, we can see
at once that there are no great correlations induced by this
simplification---in fact, the two largest absolute correlations
between the simple components are about $0.2$, for $(\hat{\alpha}%
_{1},\hat{\alpha}_{5})$ and $(\hat{\alpha}_{1},\hat{\alpha}_{4}%
)$, respectively. Overall, the scatterplot matrix is visually close to
the one given by an exact principal component analysis, but much more
interpretable.

Skipping $\hat{S}_{2}$, the two best one star solutions, $\hat{S}_{3}$
and $\hat{S}_{4}$, again both of type~\textbf{A}, are shown in
Table \ref{one stars solutions books data}. They remain clearly
interpretable, having greater overall simplicity than $\hat{S}_{1}$ and
comparable accuracies to it (indeed, $\hat{S}_{4}$ dominates
$\hat{S}_{1}$). In particular, the first two simple components comprise
an overall mean and an open-closed book contrast, $\hat{S}_{3}$ using
a weighted mean and $\hat {S}_{4}$ a weighted contrast. Between them,
our automated bias toward simplicity puts $\hat{S}_{3}$ first. In it,
all other contrasts are either within closed-book exams
($\hat{\mathbf{z}}_{3}$) or within open-book exams
($\hat{\mathbf{z}}_{4}$ and $\hat{\mathbf{z}}_{5}$). Overall,
$\hat{S}_{3}$ and $\hat{S}_{4}$ provide helpful, alternative views of
the same data. As noted above (Section \ref{sec:rg}),
\citet{RousGass2004} will not report such one star
solutions.

The default version of \citet{RousGass2004} estimates one block to be
appropriate for this data set. Our $\hat{S}_{1}$ solution coincides
with their corresponding optimal $b=1$ fit, providing empirical support
for its implicit model (see Section~\ref{sec:rg}). Although orthogonal,
their optimal $b=2$ fit does not appear among our solutions, adding
further empirical evidence that a two block model is not appropriate
for these data.

The method of \citet{Vine2000} with associated parameter $c=0$ produces
the same first and third components as our $\hat{S}_{1}$. Its other
components differ and are somewhat harder to interpret, the highest
complexity (11 for component 4) exceeding
\mbox{$N^{\ast}=9$.}

\section{Further examples} \label{sec:extra.ex}

\subsection{Reflexes data} \label{sec:reflex}

The reflexes data, taken from Section 3.8.1 of Jolliffe~(\citeyear{Joll2002}),
comprise measurements on 143 individuals of left and right reflexes for
five parts of the body, three in the upper limb and two in the lower.

\begin{table}
\caption{Exact PCA loadings (rounded to 2 decimal places) for the
reflexes data} \label{tb:reflex.ex}
\begin{tabular*}{\textwidth}{@{\extracolsep{\fill}}ld{2.2}d{2.2}d{2.2}d{1.2}d{1.2}d{1.2}d{1.2}d{1.2}d{1.2}d{1.2}@{}}
\hline
\multicolumn{1}{@{}l}{\textbf{Variable}} & \multicolumn{1}{c}{$\mathbf{q_{1}}$} &
\multicolumn{1}{c}{$\mathbf{q_{2}}$} & \multicolumn{1}{c}{$\mathbf{q_{3}}$} & \multicolumn{1}{c}{$\mathbf{q_{4}}$} &
\multicolumn{1}{c}{$\mathbf{q_{5}}$} & \multicolumn{1}{c}{$\mathbf{q_{6}}$} & \multicolumn{1}{c}{$\mathbf{q_{7}}$} &
\multicolumn{1}{c}{$\mathbf{q_{8}}$} & \multicolumn{1}{c}{$\mathbf{q_{9}}$} & \multicolumn{1}{c@{}}{$\mathbf{q_{10}}$}\\
\hline
triceps.R & 0.35 & -0.18 & 0.18 & 0.49 & -0.27 & -0.06 & -0.05 & 0.00 &
0.10 & 0.69\\
triceps.L & 0.36 & -0.19 & 0.15 & 0.47 & -0.27 & -0.02 & -0.13 & 0.01 &
-0.13 & -0.70\\
biceps.R & 0.36 & -0.13 & -0.14 & 0.04 & 0.71 &
-0.50 & -0.22 & -0.03 &
-0.19 & 0.04\\
biceps.L & 0.39 & -0.14 & -0.09 & 0.05 & 0.41 & 0.70 & 0.35 & 0.02 &
0.19 & -0.03\\
wrist.R & 0.34 & -0.24 & 0.14 & -0.51 & -0.16 &
-0.21 & -0.13 & -0.01 & 0.67 &
-0.10\\
wrist.L & 0.34 & -0.22 & 0.17 & -0.52 & -0.23 & 0.11 & 0.08 & 0.03 &
-0.67 & 0.12\\
knee.R & 0.30 & 0.29 & -0.50 & 0.02 & -0.24 &
-0.35 & 0.62 & -0.02 & 0.01 &
-0.04\\
knee.L & 0.27 & 0.35 & -0.54 & -0.07 & -0.18 & 0.28 & -0.63 & 0.02 &
-0.02 & 0.06\\
ankle.R & 0.20 & 0.53 & 0.41 & -0.03 & 0.07 & 0.03
& 0.00 & -0.71 & -0.01 &
-0.02\\
ankle.L & 0.19 & 0.54 & 0.40 & -0.02 & 0.10 & -0.04 & 0.01 & 0.70 &
0.03 & -0.01\\ [3pt]
  Variance (\%) & 52.23 & 20.36 & 10.94 & 8.57 &
4.96 & 1.08 & 0.86 & 0.59 & 0.23 & 0.19\\
\hline
\end{tabular*}
\end{table}

A principal component analysis of the correlation matrix is reported in
Table \ref{tb:reflex.ex}. This brings out some of the structure in the
data. It also provides a further example of the appropriateness of
taking equal scientific interest in all the components.

The dominant component is an overall mean, while components 2--5
contrast reflexes in different parts of the body. Smaller components
mainly contrast reflexes on the left and right sides of the body, the
substantially smaller variances associated with them suggesting near
constant linear relationships. However, more detailed interpretation of
the principal components is not immediate. For example, interpretation
of the first principal component is impaired by variability in the
loadings, notably the relatively small ones allocated to the two ankle
measurements.

\subsubsection{Results of our approach}

\begin{figure}[b]

\includegraphics{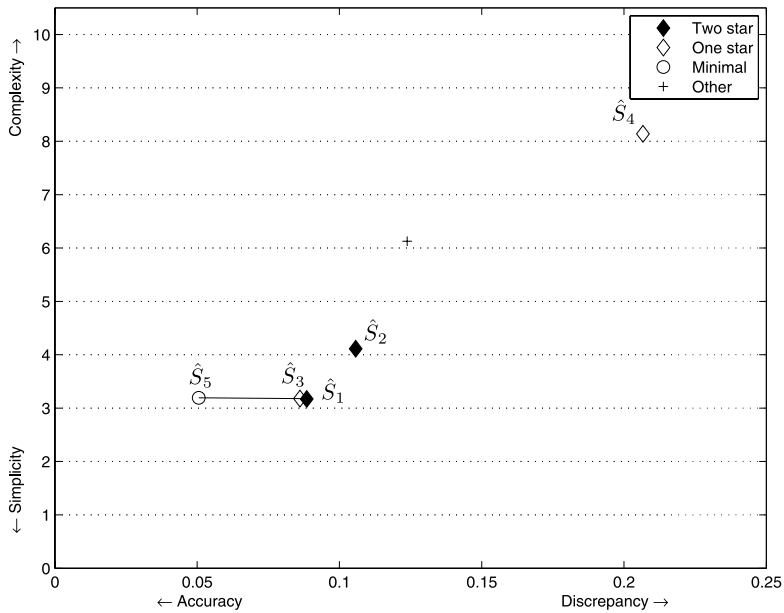}

\caption{Solution set $\mathcal{S}$ for the reflexes data.}\label{fig:reflex}
\end{figure}

Our approach provides six different solutions for these data. The
corresponding accuracy--simplicity plot (Figure \ref{fig:reflex}) shows
$\hat{S}_{1}$ and $\hat{S}_{2}$ with two stars, $\hat{S}_{3}$ and
$\hat{S}_{4} $ with one star, $\hat{S}_{5}$ as an unstarred minimal
solution, and one unlabeled `other' solution $\hat{S}_{6}$.

The user is referred first to solution $\hat{S}_{1}$, shown in
Table \ref{tb:reflex.2star}, which has the following clear
interpretation. The dominant simple component $\hat{\alpha}_{1}$\
is just the simple average of all the reflexes, while
$\hat{\alpha}_{2}$ contrasts those in upper and lower limbs.
Again, $\hat{\alpha}_{3}$ contrasts the two lower limb parts, while
$\hat{\alpha}_{4}$ and $\hat{\alpha}_{5}$ contrast the three
upper limb parts: first, triceps with wrist; then, biceps with these
two. Taking the near constant simple components in reverse order,
$\hat{\alpha}_{10}$ to $\hat{\alpha}_{8}$ suggest left--right
symmetry in triceps, wrist and ankle respectively. Finally, taking
$\hat{\alpha}_{7}$ and $\hat{\alpha}_{6}$ together as we may
(they have essentially the same variance), the two-dimensional subspace
which they span suggests left--right symmetry in knees and biceps. This
follows at once from considering their sum and difference,
corresponding to a $45^{\circ}$ rotation of axes within this subspace.

The variance explained by the first five simple components
here is $96.7\%$ compared to $97.1\%$ for the exact principal
components, each being close to that of its optimal counterpart. There
are three absolute correlations of about $0.3$ [for $ (
\hat{\alpha}_{5},\hat{\alpha}_{7} )  $, $ (\hat{\alpha}_{6},\hat{\alpha}_{9} )  $ and $ (
\hat{\alpha}_{7},\hat{\alpha}_{9} ) $], all others being
appreciably smaller.
\begin{table}\tablewidth=330pt
\tabcolsep=0pt
\caption{Integer representations for $\hat{S}_{1}$}\label{tb:reflex.2star}
\begin{tabular*}{330pt}{@{\extracolsep{\fill}}lk{2.2}k{2.2}k{2.2}k{1.2}k{1.2}k{1.2}k{1.2}k{1.3}k{1.2}k{1.2}@{}}
\hline
\multicolumn{1}{@{}l}{\textbf{Variable}} & \multicolumn{1}{c}{$\bolds{\hat{\mathbf{z}}_{1}}$} &
\multicolumn{1}{c}{$\bolds{\hat{\mathbf{z}}_{2}}$} & \multicolumn{1}{c}{$\bolds{\hat{\mathbf{z}}_{3}}$} &
\multicolumn{1}{c}{$\bolds{\hat{\mathbf{z}}_{4}}$} & \multicolumn{1}{c}{$\bolds{\hat{\mathbf{z}}_{5}}$} &
\multicolumn{1}{c}{$\bolds{\hat{\mathbf{z}}_{6}}$} & \multicolumn{1}{c}{$\bolds{\hat{\mathbf{z}}_{7}}$} &
\multicolumn{1}{c}{$\bolds{\hat{\mathbf{z}}_{8}}$} &
\multicolumn{1}{c}{$\bolds{\hat{\mathbf{z}}_{9}}$}& \multicolumn{1}{c}{$\bolds{\hat{\mathbf{z}}_{10}}$}\\
\hline
triceps.R & 1 & 2 &  & 1 & 1 &  &  &  &  & -1\\
triceps.L & 1 & 2 &  & 1 & 1 &  &  &  &  & 1\\
biceps.R & 1 & 2 &  &  & -2 & -1 & 1 &  &  & \\
biceps.L & 1 & 2 &  &  & -2 & 1 & -1 &  &  & \\
wrist.R & 1 & 2 &  & -1 & 1 &  &  &  & -1 & \\
wrist.L & 1 & 2 &  & -1 & 1 &  &  &  & 1 & \\
knee.R & 1 & -3 & -1 &  &  & -1 & -1 &  &  & \\
knee.L & 1 & -3 & -1 &  &  & 1 & 1 &  &  & \\
ankle.R & 1 & -3 & 1 &  &  &  &  & -1 &  & \\
ankle.L & 1 & -3 & 1 &  &  &  &  & 1 &  & \\ [6pt]
Accuracy & \multicolumn{1}{d{2.2}}{0.98} & \multicolumn{1}{d{2.2}}{0.95} & \multicolumn{1}{d{2.2}}{0.92}
& \multicolumn{1}{d{1.2}}{0.99} & \multicolumn{1}{d{1.2}}{0.91} & \multicolumn{1}{d{1.2}}{0.91}
& \multicolumn{1}{d{1.2}}{0.91} & \multicolumn{1}{d{1.3}}{0.998} & \multicolumn{1}{d{1.2}}{0.95}
& \multicolumn{1}{d{1.2}@{}}{0.98}\\ [3pt]
 Variance (\%) & \multicolumn{1}{d{2.2}}{50.8} & \multicolumn{1}{d{2.2}}{20.6} & \multicolumn{1}{d{2.2}}{11.2}
& \multicolumn{1}{d{1.2}}{8.6} & \multicolumn{1}{d{1.2}}{5.6} & \multicolumn{1}{d{1.2}}{1.1}
& \multicolumn{1}{d{1.2}}{1.1} & \multicolumn{1}{d{1.3}}{0.6} & \multicolumn{1}{d{1.2}}{0.3} &
\multicolumn{1}{d{1.2}@{}}{0.2}\\
\hline
\end{tabular*}\vspace*{-6pt}
\legend{\textit{Notes:} Reflexes data. Empty entries mean zeroes.}
\end{table}

The one star solution $\hat{S}_{3}$ is very close to $\hat{S}_{1}$ in
Figure \ref{fig:reflex}. Indeed, it differs from it only in
$\hat{\alpha}_{6}$ and $\hat{\alpha}_{7}$ representing another rotation
within their span, already interpreted above as suggesting left--right
symmetry in knees and biceps. This time, at the cost of increasing the
complexity of both $\hat{\alpha}_{6} $ and $\hat{\alpha}_{7}$ by one,
their accuracies improve to $0.95$ and $0.97$, respectively. This
illustrates that subspace rotation can increase accuracy without
changing overall interpretation.

Although dominated by $\hat{S}_{1}$, the other two star solution
$\hat{S}_{2} $ provides an interesting alternative view. It differs
only on two components, both having very clear\
interpretations:

\begin{longlist}
\item[$\hat{\alpha}_{2}$:] Contrasts upper and lower limbs, omitting
biceps, using only $\pm1$ loadings.

\item[$\hat{\alpha}_{5}$:] Contrasts biceps with everything else,
using only 1 and $-4$ loadings.
\end{longlist}

The other, unstarred, minimal solution $\hat{S}_{5}$ (details not
shown) is simple and accurate, but somewhat less clearly structured.
Its $\hat{\alpha }_{4}$ and $\hat{\alpha}_{8}$ to $\hat{\alpha}_{10}$
agree exactly with $\hat{S}_{1}$, the sign pattern of
$\hat{\alpha}_{6}$ and $\hat{\alpha}_{7}$ also agreeing (each now has
loadings of $\pm2$). $\hat{\alpha}_{1}$ is a weighted mean, omitting
ankle. $\hat{\alpha}_{5}$ also omits ankle,
contrasting biceps with the other three body parts. $\hat{\alpha}_{2}%
$ contrasts upper and lower limbs, omitting biceps. Finally,
$\hat{\alpha }_{3}$ also omit biceps, contrasting knee with the other
three body parts.

The other two solutions, $\hat{S}_{4}$ and $\hat{S}_{6}$, are markedly
less simple and accurate.

\subsubsection{Comparison with other approaches}\label{sec:reflex.comp}

We briefly compare our results here with those of other methods.

The comparison with Rousson and Gasser's approach for these data is,
essentially, the same as it was for the running exams data (see the end
of Section \ref{sec:exams.res}). The default version of
\citet{RousGass2004} again estimates $b=1$, our $\hat{S}_{1}$ solution
coinciding with their corresponding optimal fit, providing empirical
support for its implicit model. Although orthogonal, their optimal
$b=2$ fit does not appear among our solutions, adding further
empirical evidence that a two block model is not appropriate for these
data.

\begin{table}[b]
\tabcolsep=0pt
\caption{Integer representations for the reflexes data using Vines'
method with $c=0$} \label{Reflex Vines
Method}
{\fontsize{8.3}{10}\selectfont{
\begin{tabular*}{\textwidth}{@{\extracolsep{\fill}}lk{2.2}k{2.2}k{1.2}k{1.2}k{1.2}k{1.2}k{1.2}k{1.2}k{1.2}k{1.2}@{}}
\hline
\multicolumn{1}{@{}l}{\textbf{Variable}} & \multicolumn{1}{c}{\hspace*{1.8pt}$\bolds{\hat{\mathbf{z}}_{1}}$} & \multicolumn{1}{c}{\hspace*{0.5pt}$\bolds{\hat{\mathbf{z}}_{2}}$}
& \multicolumn{1}{c}{\hspace*{-4pt}$\bolds{\hat{\mathbf{z}}_{3}}$} & \multicolumn{1}{c}{\hspace*{-4pt}$\bolds{\hat{\mathbf{z}}_{4}}$} & \multicolumn{1}{c}{\hspace*{-9pt}$\bolds{\hat{\mathbf{z}}_{5}}$}
& \multicolumn{1}{c}{\hspace*{-17pt}$\bolds{\hat{\mathbf{z}}_{6}}$} & \multicolumn{1}{c}{\hspace*{-16pt}$\bolds{\hat{\mathbf{z}}_{7}}$} & \multicolumn{1}{c}{\hspace*{-15.5pt}$\bolds{\hat{\mathbf{z}}_{8}}$}
& \multicolumn{1}{c}{\hspace*{-15.5pt}$\bolds{\hat{\mathbf{z}}_{9}}$} & \multicolumn{1}{c}{\hspace*{-4pt}$\bolds{\hat{\mathbf{z}}_{10}}$}\\
\hline
triceps.R & 1. & 1. & 2. & 1. & 19. & -19. & 19. & 19. & -19. & -1\\
triceps.L & 1. & 1. & 2. & 1. & 19. & -19. & 19. & 19. & -19. & 1\\
biceps.R & 1. & 1. & -1. &  & -42. & -2479. & -42. & -42. & 42. & \\
biceps.L & 1. & 1. & -1. &  & -40. & 2561. & -40. & -40. & 40. & \\
wrist.R & 1. & 1. & 2. & -1. & 18. & -19. & 19. & 18. & -2541. & \\
wrist.L & 1. & 1. & 2. & -1. & 20. & -19. & 19. & 20. & 2503. & \\
knee.R & 1. & -1. & -9. &  & 10. & -9. & -2512. & 10. & -10. & \\
knee.L & 1. & -1. & -9. &  & 8. & -9. & 2530. & 8. & -8. & \\
ankle.R & 1. & -2. & 6. &  & -5. & 6. & -6. & -2529. & 6. & \\
ankle.L & 1. & -2. & 6. &  & -7. & 6. & -6. & 2517. & 6. & \\ [6pt]
Accuracy & \multicolumn{1}{d{2.2}}{0.98} & \multicolumn{1}{d{2.2}}{0.97} & \multicolumn{1}{d{1.2}}{0.99}
& \multicolumn{1}{d{1.2}}{0.99} & \multicolumn{1}{d{1.2}}{0.97} & \multicolumn{1}{d{1.2}}{0.85} &
\multicolumn{1}{d{1.2}}{0.89} & \multicolumn{1}{d{1.2}}{1.00} &
\multicolumn{1}{d{1.2}}{0.95}
& \multicolumn{1}{d{1.2}}{0.98}\\ [3pt]
Variance (\%)  & \multicolumn{1}{d{2.2}}{50.8} & \multicolumn{1}{d{2.2}}{21.5} & 11 &
\multicolumn{1}{d{1.2}}{8.6} & 5 & \multicolumn{1}{d{1.2}}{1.2} & \multicolumn{1}{d{1.2}}{0.99} &
\multicolumn{1}{d{1.2}}{0.60} & \multicolumn{1}{d{1.2}}{0.30} & \multicolumn{1}{d{1.2}}{0.20}\\
\hline
\end{tabular*}\vspace*{-6pt}
}}
\legend{\textit{Note:} Empty entries mean zeroes.}
\end{table}

Table \ref{Reflex Vines Method} shows the components obtained using the
method of  \citet{Vine2000}  with associated parameter $c=0$. Compared
to the original principal component analysis
(Table \ref{tb:reflex.ex}), this gives a substantially simpler, more
interpretable solution. It differs from $\hat{S}_{1}$, especially for
middle components, but interestingly picks up
the same simplified components $\hat{\alpha}_{1}$, $\hat{\alpha}_{4}%
$ and $\hat{\alpha}_{10}$, interpreted above (this might, in part,
be because Vines' method is able to seek simplifications of components
in a nonsequential fashion). Axes $\hat{\alpha}_{8}$ and
$\widehat{\alpha }_{9}$ here have virtually the same accuracy as in
$\hat{S}_{1}$ but are much more complex, illustrating that our bias
toward simplicity does not necessarily sacrifice accuracy. Indeed,
having a dominant pair of elements of nearly equal size and opposite
sign, $\hat{\alpha}_{8}$ and $\hat{\alpha}_{9}$ are both
angle-close to the corresponding axes in $\hat{S}_{1}$, interpreted
above as suggestive of left--right symmetry in the corresponding part of
the body. By the same token, $\hat{\alpha}_{6}$ and
$\hat{\alpha}_{7}$ are also angle-close to suggesting
corresponding left--right symmetries. The remaining axes,
$\hat{\alpha}_{2}$, $\hat{\alpha}_{3}$ and
$\hat{\alpha}_{5}$, are more accurate, but less directly
interpretable, than those in $\hat{S}_{1}$.

\begin{table}\tablewidth=280pt
\caption{Exact principal component analysis loadings (rounded to 2
decimal places) for the alate adelges data} \label{alate adelges data
exact PCA}
\begin{tabular*}{280pt}{@{\extracolsep{\fill}}ld{2.2}d{2.2}d{1.2}d{1.2}@{}}
\hline
\multicolumn{1}{@{}l}{\textbf{Variable}} & \multicolumn{1}{c}{$\mathbf{q_{1}}$} & \multicolumn{1}{c}{$\mathbf{q_{2}}$} &
\multicolumn{1}{c}{$\mathbf{q_{3}}$} & \multicolumn{1}{c@{}}{$\mathbf{q_{4}}$}\\
\hline
Length & 0.25 & 0.03 & 0.02 & 0.07\\
Width & 0.26 & 0.07 & 0.01 & 0.10\\
Forwing & 0.26 & 0.03 & -0.05 & 0.07\\
Hinwing & 0.26 & 0.09 & 0.03 & 0.00\\
Antseg 1 & 0.24 & -0.18 & 0.04 & -0.01\\
Antseg 2 & 0.25 & -0.16 & 0.00 & 0.02\\
Antseg 3 & 0.23 & 0.24 & 0.05 & 0.11\\
Antseg 4 & 0.24 & 0.04 & 0.16 & 0.01\\
Antseg 5 & 0.25 & -0.03 & 0.10 & -0.02\\
Tarsus 3 & 0.26 & 0.01 & 0.03 & 0.18\\
Tibia 3 & 0.26 & 0.03 & 0.08 & 0.20\\
Femur 3 & 0.26 & 0.07 & 0.12 & 0.19\\
Rostrum & 0.25 & -0.01 & 0.07 & 0.04\\
Ovipositor & 0.20 & -0.40 & -0.02 & 0.06\\
Spiracles & 0.16 & -0.41 & -0.19 & -0.62\\
Ov-spines & 0.11 & -0.55 & -0.15 & 0.04\\
Anal fold & -0.19 & -0.35 & 0.04 & 0.49\\
Ant-spines & -0.13 & -0.20 & 0.93 & -0.17\\
Hooks & 0.20 & 0.28 & 0.05 & -0.45\\ [3pt]
Variance (\%)  & 73.0 & 12.5 &
3.9 & 2.6\\
\hline
\end{tabular*}
\end{table}

\subsection{Alate adelges data}\label{sec:aphids}

These data consist of 19 anatomical measurements of~40 \textit{alate
adelges} (winged aphids), as reported in \citet{Jeff1967}. The
measurements taken on each aphid are its length and width, fore-wing
and hind-wing lengths, 5 antennal segment lengths, 3 leg bone
measurements, measurements of the rostrum and the ovipositor, anal
fold, and counts of the number of spiracles, ovipositor spines,
antennal spines and hind-wing hooks.

\citet{Jeff1967} focuses attention on the $k=4$ dominant eigenvectors
of the correlation matrix shown in Table \ref{alate adelges data exact
PCA}, these accounting for $92\%$ of the total variability in the data.
He interprets $\alpha_{1}$ as a general index of size, and
$\alpha_{2}$ to $\alpha_{4}$ as essentially measuring the number of
ovipositor spines, of antennal spines and of spiracles, respectively.

These tidy interpretations are not without difficulty. For
$\alpha_{1}$, some later variables have notably smaller loadings, of
both signs. For each other $\alpha_{r}$, the interpretation offered
amounts to `thresholding' (setting all smaller loadings to zero) at the
maximum absolute value of $q_{r}$. Whereas this looks quite reasonable
for $\alpha_{3}$, it seems much less so for $\alpha_{2}$ and
$\alpha_{4}$, these axes containing a range of substantial loadings,
some of comparable magnitude to their maximum.

Inspection of the correlation matrix in \citet{Jeff1967} shows that,
while positively correlated with each other, the two variables with
negative loadings on $\alpha_{1}$ are, with one insignificant
exception, negatively correlated with all the other variables. Indeed,
a single negative (at $-0.026$, essentially zero) correlation remains
when both their signs are reversed. Following \citet{RousGass2003},
one strategy is to reverse these two signs, analyze the data in some
way and then, to retain the interpretation of the original variables,
switch them back again. We call this process `sign reversal.'

We compare here three solutions for these data, detailed in
Table \ref{alate adelges_nextbest_forwards_best and RG}:

\begin{itemize}
\item $\hat{S}_{1}$, as defined above,

\item $\widetilde{S}_{1}$, the result of an $\hat{S}_{1}$ analysis
with sign reversal, and

\item $\widetilde{S}_{\mathit{RG}}$, an optimal \citet{RousGass2004} fit with
$b=1$ and, again, sign reversal.
\end{itemize}
Vines' method is not capable to produce any answer here,
mainly due to the complexity of some of the approximate loading vectors
growing far too big.

Whereas none is ideal (in particular, there is substantial correlation
in each, especially $\hat{S}_{1}$), these three solutions provide
helpful, complementary views of these data. We discuss them in turn.

\begin{table}
\tabcolsep=0pt
\caption{Integer representations for the alate adelges data}\label{alate adelges_nextbest_forwards_best and RG}
\begin{tabular*}{\textwidth}{@{\extracolsep{\fill}}lk{2.2}k{1.2}k{1.2}k{1.2}k{2.2}k{2.2}k{1.2}k{1.2}k{2.2}k{2.2}k{1.2}k{1.2}@{}}
\hline
&\multicolumn{4}{c}{$\bolds{\hat{S}_{1}}$} &
\multicolumn{4}{c}{$\bolds{\widetilde{S}_{1}}$} &
\multicolumn{4}{c@{}}{$\bolds{\widetilde {S}_{\mathit{RG}}}$}\\ [-7pt]
&\multicolumn{4}{@{}c}{\hrulefill}&\multicolumn{4}{c}{\hrulefill}&\multicolumn{4}{c@{}}{\hrulefill}\\
\multicolumn{1}{@{}l}{\textbf{Variable}} & \multicolumn{1}{c}{$\bolds{\hat{\mathbf{z}}_{1}}$}
& \multicolumn{1}{c}{$\bolds{\hat{\mathbf{z}}_{2}}$} & \multicolumn{1}{c}{$\bolds{\hat{\mathbf{z}}_{3}}$} &
\multicolumn{1}{c}{$\bolds{\hat{\mathbf{z}}_{4}}$} & \multicolumn{1}{c}{$\bolds{\hat{\mathbf{z}}_{1}}$} &
\multicolumn{1}{c}{$\bolds{\hat{\mathbf{z}}_{2}}$} & \multicolumn{1}{c}{$\bolds{\hat{\mathbf{z}}_{3}}$} &
\multicolumn{1}{c}{$\bolds{\hat{\mathbf{z}}_{4}}$} & \multicolumn{1}{c}{$\bolds{\hat{\mathbf{z}}_{1}}$} &
\multicolumn{1}{c}{$\bolds{\hat{\mathbf{z}}_{2}}$} & \multicolumn{1}{c}{$\bolds{\hat{\mathbf{z}}_{3}}$} &
\multicolumn{1}{c@{}}{$\bolds{\hat{\mathbf{z}}_{4}}$}\\
\hline
Length & 1 &  &  &  & 2 &  &  &  & 1 &  & 3 & \\
Width & 1 &  &  &  & 2 &  &  &  & 1 &  & 3 & 1\\
Forwing & 1 &  &  &  & 2 &  &  &  & 1 &  &  & 1\\
Hinwing & 1 &  &  &  & 2 &  &  &  & 1 &  & 3 & \\
Antseg 1 & 1 &  &  &  & 2 & 1 &  &  & 1 &  &  & \\
Antseg 2 & 1 &  &  &  & 2 & 1 &  &  & 1 &  &  & \\
Antseg 3 & 1 & -1 &  &  & 2 & -1 &  &  & 1 & 3 & 3 & 1\\
Antseg 4 & 1 &  &  &  & 2 &  &  &  & 1 &  & 3 & \\
Antseg 5 & 1 &  &  &  & 2 &  &  &  & 1 &  & 3 & \\
Tarsus 3 & 1 &  &  &  & 2 & -1 &  & 1 & 1 &  & 3 & 1\\
Tibia 3 & 1 &  &  &  & 2 &  &  & 1 & 1 &  & 3 & 1\\
Femur 3 & 1 &  &  &  & 2 &  &  & 1 & 1 &  & 3 & 1\\
Rostrum & 1 &  &  &  & 2 &  &  &  & 1 &  & 3 & \\
Ovipositor & 1 & 1 &  &  & 1 & 2 &  &  & 1 & -4 &  & 1\\
Spiracles &  & 1 &  & 1 &  & 2 & 1 & -2 & 1 & -4 & -11 & -3\\
Ov-spines &  & 1 &  &  & 1 & 2 &  &  & 1 & -4 & -11 & 1\\
Anal fold &  & 1 &  & -1 & -1 & 2 &  & 2 & -1 & -3 &  & 3\\
Ant-spines &  &  & 1 &  &  & 1 & -2 & -1 & -1 & -3 & 11 & -1\\
Hooks &  &  &  & 1 & 2 & -1 &  & -2 & 1 & 3 & 3 & -3\\ [6pt]
Accuracy &
\multicolumn{1}{d{2.2}}{0.93} & \multicolumn{1}{d{1.2}}{0.87} & \multicolumn{1}{d{1.2}}{0.93}
& \multicolumn{1}{d{1.2}}{0.90} & \multicolumn{1}{d{2.2}}{0.97} & \multicolumn{1}{d{2.2}}{0.95}
& \multicolumn{1}{d{1.2}}{0.92} & \multicolumn{1}{d{1.2}}{0.96} & \multicolumn{1}{d{2.2}}{0.98}
& \multicolumn{1}{d{2.2}}{0.94} &\multicolumn{1}{d{1.2}}{0.75} &
\multicolumn{1}{d{1.2}}{0.96}\\ [3pt]
 Variance (\%)& \multicolumn{1}{d{2.2}}{63.5} & \multicolumn{1}{d{1.2}}{9.7} & \multicolumn{1}{d{1.2}}{5.3}
& \multicolumn{1}{d{1.2}}{9.8} & \multicolumn{1}{k{2.2}}{69} & \multicolumn{1}{d{2.2}}{11.6} &
\multicolumn{1}{d{1.2}}{6} & \multicolumn{1}{d{1.2}}{2.7} & \multicolumn{1}{d{2.2}}{70.2} &
\multicolumn{1}{d{2.2}}{11.3} & \multicolumn{1}{d{1.2}}{7.8} & \multicolumn{1}{d{1.2}}{3}\\[-7pt]
&\multicolumn{4}{c}{\hrulefill}&\multicolumn{4}{c}{\hrulefill}&\multicolumn{4}{c@{}}{\hrulefill}\\
 Optimality (\%)
&\multicolumn{4}{d{2.2}}{86.9} & \multicolumn{4}{d{2.2}}{94.1} &
\multicolumn{4}{d{2.2}@{}}{94.5}\\
Max correl &
\multicolumn{4}{d{2.2}}{0.83} & \multicolumn{4}{d{2.2}}{0.63} &
\multicolumn{4}{d{2.2}@{}}{0.63}\\
\hline
\end{tabular*}\vspace*{-6pt}
\legend{\textit{Note:} Empty entries mean zeroes.}
\end{table}

As expected, given that $\alpha_{1}$ is not single-signed,
$\hat{S}_{1}$ is unstarred. Nevertheless, it is perhaps the most
easily interpreted solution. It is the simplest and sparsest, all
loadings being 0, 1 or $-1$. Its dominant component is the simple average
of all the variables, excluding the four count variables and anal fold.
Its third component is the number of antennal spines. The
other two components are simple contrasts, whose pattern of zeroes is
consistent with thresholding at lower levels with only two exceptions
(the last two loadings in $\hat{\alpha}_{2}$), these zeroes
ensuring orthogonality. However, $\hat{\alpha}_{2}$ is not very
accurate and, indeed, explains less variance than $\hat{\alpha}_{4}$.

$\widetilde{S}_{1}$ is the most accurate solution, the minimum accuracy
being 0.92. Although also unstarred, it is perhaps the next most easily
interpreted. It is nearly as simple and as sparse as $\hat{S}_{1}$. It
has a comparable corrected sum of variances to the optimized
$\widetilde{S}_{\mathit{RG}}$ fit (94.1\% compared to 94.5\%), achieved despite
having lower variances associated with the last two components,
consistent with their suggestion of underlying regularities. Compared
to $\hat{S}_{1}$, its dominant component gains accuracy and variance
explained, but is less easily interpreted. Finally, the pattern of
zeroes in its other components is consistent with thresholding at yet
lower levels with only one exception (for Tarsus 3 on
$\hat{\alpha}_{2}$), this nonzero loading ensuring orthogonality.

A $b=1$ solution such as $\widetilde{S}_{\mathit{RG}}$ comes from fitting the
following assumed form of two star solution to the (sign reversed)
data: a simple arithmetic mean, plus a set of contrasts in each of
which the nonzero elements comprise $m$ times a value $n$, and $n$
times a value $-m$. As happens here, these contrasts need not be
orthogonal, so that $\widetilde{S}_{\mathit{RG}}$ cannot appear among our
solutions. Its dominant component fits well, having the highest
accuracy and variance explained, while the zeroes in its second
component are, without exception, consistent with the same lower
thresholding
as in $\hat{S}_{1}$. However, the other fits seem poor, $\hat{\alpha}_{3}%
$ having a particularly low accuracy, while $\hat{\alpha}_{4}$ is
considerably less sparse than in $\widetilde{S}_{1}$ but without
improving accuracy. Overall, despite dropping the orthogonality
constraint, $\widetilde{S}_{\mathit{RG}}$ comes third in terms of simplicity,
accuracy and sparseness. A likely reason for this is that its assumed
model seems, at most, appropriate to the first two components.

\subsection{Larger data sets}\label{sec:bigp}

In this section we use simulated examples to give an idea of how our
method behaves when the number of variables $p$ grows. These examples
also illustrate the secondary, initially surprising, fact that certain
simple structures in the population principal components can be
recovered using only information from the sample. Such behavior has
been observed, for small dimensions, in one other simplification
method: see \citet{SunL2006}.

For $p=$ 8, 16, 32, 64, 128 and 256, we simulated 100 data sets of size
$n$ from a $p$-variate, zero mean, normal distribution with covariance
matrix of the following form. Its matrix $\mathbf{Q}_{\mathrm{pop}}$ of
population eigenvectors is the particular integer matrix with
orthogonal columns $\mathbf{Z}_{\mathrm{pop}}$ detailed below, normalized to
unit column length. Its spectrum has four reasonably well-separated
dominant eigenvalues $(16,8,4,2)$, the rest being equal with sum $1$.
Thus, for each $p$, the first four population components explain
$30/31\sim97\%$ of total variability, the corresponding four sample
components being used as input data here in each case. Sampling
variability was kept constant across different values of $p$ in the
sense that the ratio of the number of degrees of freedom in the
centered data to that in $\mathbf{Q}_{\mathrm{pop}}$ was kept fixed at $8$,
giving $n=4p-3$.

We use the following two star structure for the population eigenaxes
generated by $\mathbf{Z}_{\mathrm{pop}}$. A so-called Hadamard matrix of order
$p=2^{m}$ can be
obtained inductively using%
\[
\mathbf{Z}_{\mathrm{pop}}(2)=\pmatrix{
1 & 1\cr
1 & -1
}
\]
and
\[
\mbox{for }m>1\qquad\mathbf{Z}_{\mathrm{pop}}(2^{m})=\pmatrix{
\mathbf{Z}_{\mathrm{pop}}(2^{m-1}) & \mathbf{Z}_{\mathrm{pop}}(2^{m-1})\cr
\mathbf{Z}_{\mathrm{pop}}(2^{m-1}) & -\mathbf{Z}_{\mathrm{pop}}(2^{m-1})
}.
\]
For example, this gives%
\[
\mathbf{Z}_{\mathrm{pop}}(4)= \pmatrix{
1 & 1 & 1 & 1\cr
1 & -1 & 1 & -1\cr
1 & 1 & -1 & -1\cr
1 & -1 & -1 & 1
}
\]
(axis-equivalent to that found in the blood flow data of Section
\ref{sec:ri}). This structure is the opposite of sparse, having no
zeroes. Instead, it has what  \citet{ChipGu2005} call `homogeneity.'
At the same time, it is extremely simple. For any $m$, the $\lambda$
part of our overall complexity measure (Section~\ref{sec:accsimp.plot})
takes its maximum value $1/2$, so that $\mathit{compl}(S_{\mathrm{pop}})=1.5$ if and only
if $\mathbf{Z}_{\mathrm{pop}}$ has this Hadamard form.

\begin{figure}

\includegraphics{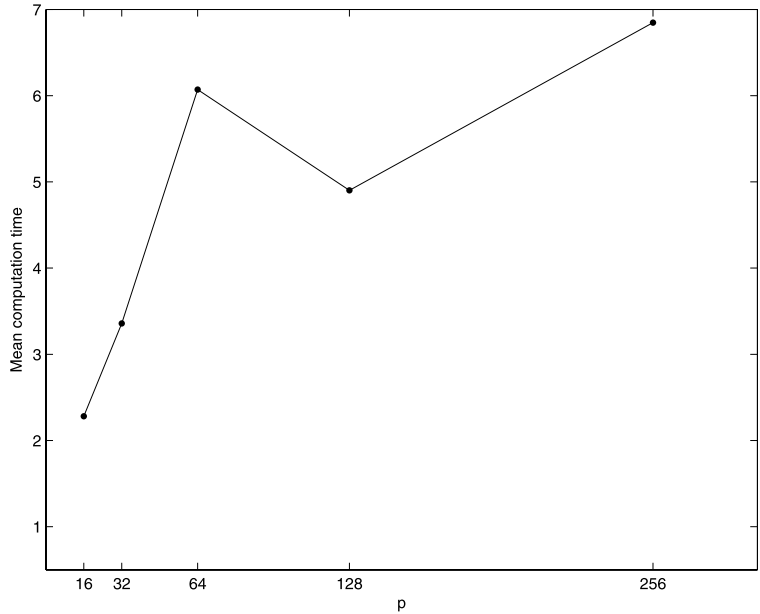}

\caption{Mean computation times relative to the time for $p=8$.}\label{fig:timing}
\end{figure}

Figure \ref{fig:timing} shows that the computation time required grows
roughly linearly in $p$, which gives a good indication that the method
is relatively quick when $p\leq256$ and there is a simple structure in
the sample eigenvectors.

\begin{figure}

\includegraphics{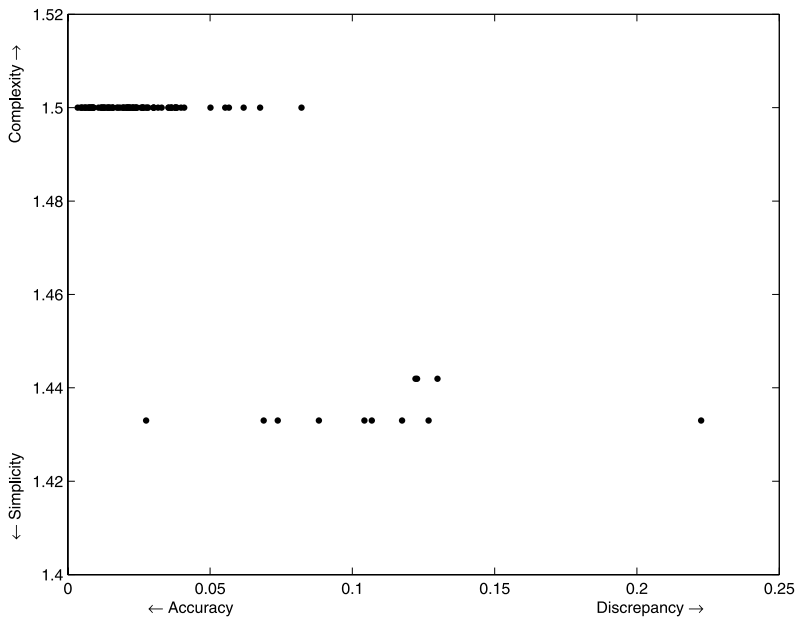}

\caption{Accuracy and complexity for solution to simulated data when
$p=32$.} \label{fig:p32}
\end{figure}

Figure \ref{fig:p32} is an accuracy--simplicity scatterplot for the
$100$ simulated values of $\hat{S}_{1}$ obtained with $p=32$. The
percentage\vspace*{1pt} of simulations with $\mathit{compl}(\hat{S}_{1})=1.5$, corresponding
to $\hat{S}_{1}$ having a Hadamard structure, is substantial. Overall,
this percentage was found to increase with $p$, as was the minimum
accuracy attained.

%

\section{Discussion}\label{sec:discuss}

Combining principles with pragmatism, a new approach and accompanying
algorithm to interpret (a subset of) principal components have been
presented and shown to work well on a range of examples. The key idea
is to approximate each eigenvector involved by an integer vector close
to it in angle terms, while keeping the size of its maximum element as
low as possible. Requiring orthogonality, attractive visualization and
dimension reduction features of principal component analysis are
retained. Being essentially exploratory, alternative views of the same
data are provided in a clear, principled order. The user is then free
to choose the set of solutions that best match his or her trade-off
between simplicity and accuracy. Again, other things being equal,
explicit models can be checked by seeing if their fits occur in our
exploratory analysis (Sections \ref{sec:exams.res} and
\ref{sec:reflex.comp}), while alternatives can be provided where
preconceived models appear inappropriate (Section \ref{sec:aphids}).
Although not directly targeted, sparsity can emerge where appropriate,
as in the example in each of the three Sections just cited.
Section \ref{sec:bigp} gives some idea of our algorithm's performance
in larger data sets, while also illustrating that sparsity is not
always appropriate. Overall, this new tool adds to the applied
statistician's armoury, effectively combining simplicity, retention of
optimality and computational efficiency, while complementing existing
methods.

Although the examples given establish that our approach is useful in
practice, an extensive simulation study is required to more fully
explore its performance and to compare it with other simplification
methods, such as those proposed by Rousson and Gasser, Chipman and Gu,
and Vines. Such a simulation study would also provide further
information about appropriate default values for the tuning parameters
employed and help to identify possible alternative measures of
interpretability, simplicity and accuracy that both highlight the best
solutions and most effectively indicate situations where simple
structures are perhaps not there to be found.

Any approach to interpreting principal components involves making
specific choices and an overall compromise between conflicting
objectives. Variants and extensions of the approach presented here
meriting future study include:

\begin{itemize}
\item exploring the potential usefulness of sequences of approximation
other than the four employed here (Section \ref{sec:seq.app}); more
radical is the possibility of simplifying two or higher-dimensional
subspaces of eigenvectors at each step;

\item varying the minimum accuracy required across eigenaxes, for
example, to reflect situations where it is more important for some
components to be approximated accurately than others (in particular,
this may be useful in connection with the variant discussed next);

\item adapting it to reflect scientific contexts in which interest
centers solely on, say, explaining variability;

\item trading off the benefits of orthogonality against the advantages
of separately approximating each eigenaxis;

\item applying its ideas in other contexts, including Linear
Discriminant Analysis and Canonical Correlation Analysis.
\end{itemize}

\begin{appendix}

\section{Distance interpretation of the~minimum~accuracy attained}\label{app:min.dist}

We show here that the minimum accuracy attained is a known, strictly
decreasing, function of a natural measure of distance between any two
ordered sets of axes.

For any two vectors $\mathbf{x}$ and $\mathbf{y}$ in
$\mathbb{R}^{p}$ with unit length, define the angle
$0\leq\theta\leq\pi$ between them by
$\cos(\theta)=\mathbf{x}^{T}\mathbf{y}$ and the following measure of
discrepancy between the axes $\pm\mathbf{x}$ and $\pm\mathbf{y}$\
which they
generate:
\[
\delta(\pm\mathbf{x},\pm\mathbf{y}):=\min\bigl\{ \Vert \mathbf{u}%
-\mathbf{v} \Vert /\sqrt{2}\dvtx\mathbf{u\in\{x,-x\},v\in\{y,-y\}}\bigr\}.%
\]
Then, omitting the straightforward proof, we have that
\[
\delta(\pm\mathbf{x},\pm\mathbf{y})=\min\bigl\{ \Vert \mathbf{x}-\mathbf{y}%
 \Vert /\sqrt{2}, \Vert \mathbf{x}+\mathbf{y} \Vert
/\sqrt {2}\bigr\}=\sqrt{1- \vert \cos(\theta) \vert }
\]
is a distance function on the set of all axes in $\mathbb{R}^{p}$
(i.e., is nonnegative, zero only when
$\pm\mathbf{x}=\pm\mathbf{y}$, symmetric and obeys the triangle
inequality), the angle-accuracy attained measure being thus a strictly
decreasing function of it, namely,
\[
 \vert \cos(\theta) \vert =1-\delta^{2}(\pm\mathbf{x},\pm
\mathbf{y}).%
\]
For any two ordered sets of axes $\mathbf{\pm X:=(\pm x}%
_{1}|\cdots|\mathbf{\pm x}_{m}\mathbf{)}$ and $\mathbf{\pm Y:=(\pm y}%
_{1}|\break\cdots|\mathbf{\pm y}_{m}\mathbf{)}$ in $\mathbb{R}^{p},$ with
$ \Vert \mathbf{x}_{r} \Vert = \Vert
\mathbf{y}_{r} \Vert =1$ and corresponding angles
$0\leq\theta_{r}\leq\pi$ given by $\cos(\theta
_{r})=\mathbf{x}_{r}^{T}\mathbf{y}_{r}$ ($1\leq r\leq m$), define now
the
following overall discrepancy measure between them:
\[
\Delta(\mathbf{\pm X},\mathbf{\pm Y}):=\max\{\delta(\pm\mathbf{x}_{r}%
,\pm\mathbf{y}_{r})\dvtx 1\leq r\leq m\}.
\]
Then, using the properties of $\delta(\cdot,\cdot)$ just established,
and
again omitting the straightforward proof, we have that%
\[
\Delta(\mathbf{\pm X},\mathbf{\pm Y})=\sqrt{1-\min\{ \vert
\cos(\theta
_{r}) \vert \dvtx1\leq r\leq m\}}%
\]
is a distance function on the set of all ordered sets of axes in
$\mathbb{R}^{p}$, the minimum angle-accuracy attained measure being
thus a
strictly decreasing function of it, namely,%
\[
\min\{ \vert \cos(\theta_{r}) \vert \dvtx 1\leq r\leq m\}=1-\Delta^{2}%
(\pm\mathbf{X},\pm\mathbf{Y}).%
\]

\section{Implementation}\label{app:impl}

In Section \ref{app:nratio} we define a key approximation to the solution of
the problem of minimizing accuracy without orthogonality restrictions,
for a given complexity. In Section \ref{app:iter} we outline approaches to the
search for $\hat{\alpha}_{r}(\theta)$.

A set of \texttt{R} routines implementing our approach is available
from the authors upon request.

\subsection{$N$-ratio simplification}\label{app:nratio}

For a given vector $\mathbf{u}\in\mathbb{R}^{(l)}$ $(l\geq2)$ and given
complexity $N$, we describe here an approximation to the solution of
the problem of maximizing $\mathit{accu}(\ell(\mathbf{u}),\ell(\mathbf{z}))$
over $\mathbf{z}\in\mathbb{Z}^{(l)}$ subject to $\mathit{compl}(\mathbf{z})=N$.

A necessary condition for $\mathbf{z}$ to be optimal is that
$\mathbf{u}$ and $\mathbf{z}$ have the same signs, while the rank
vector of $|\mathbf{z}|$ coincides with that of $|\mathbf{u}|$. Thus,
subsuming sign changes and a permutation as required, there is no loss
in taking $u_{1}\geq u_{2}\geq \cdots\geq u_{l}\geq0$ and restricting
attention to integer vectors $\mathbf{z}$ such that $z_{1}\geq
z_{2}\geq\cdots\geq z_{l}\geq0$, the corresponding inverse permutation
and sign changes being applied at the end.

The $N$\textit{-ratio simplification} of $\mathbf{u}$ is defined as
$\hat{\mathbf{z}}^{(N)}=(N,\hat{z}_{2}^{(N)},\ldots,\hat{z}_{l}^{(N)})^{\top}$
in which the $\{\hat{z}_{r}^{(N)}\}_{r=2}^{l}$ are chosen so that each
$\xi_{r}:=\tan^{-1}(\hat{z}_{r}^{(N)}/N)$ is as close as possible to
$\psi _{r}:=\tan^{-1}(\lambda_{r})$ where $\lambda_{r}:=u_{r}/u_{1}$ (a
final division by $\operatorname{hcf}(|\hat{\mathbf{z}}^{(N)}|)$ being left implicit).
Explicitly, for each $r=2,\ldots,l$, defining $l_{r}$ as the integer
part of $\lambda _{r}N$ and
$0\leq\alpha_{r}\leq\psi_{r}<\beta_{r}\leq\pi/4 $ by $\alpha
_{r}:=\tan^{-1}(l_{r}/N)$ and $\beta_{r}:=\tan^{-1}((l_{r}+1)/N)$, we
put
\begin{equation}
\hat{z}_{r}^{(N)}:= \cases{
l_{r}, &\quad if $\psi_{r}\leq(\alpha_{r}+\beta_{r})/2$,\cr
l_{r}+1, &\quad if $\psi_{r}>(\alpha_{r}+\beta_{r})/2$.
}
\end{equation}
The accuracy of this approximation comes from the fact that $\ell
(\hat{\mathbf{z}}^{(N)})=\ell(\mathbf{u})$ if and only if $\hat{z}_{r}%
^{(N)}/N=u_{r}/u_{1}$ for each $r=2,\ldots,l$. This is a very fast
approximation since, reordering of elements apart, the computational
effort involved is linear in $l$.

$N$-ratio simplification has the additional advantage that
\textit{neighboring solutions} close to $\hat{\mathbf{z}}^{(N)}$ can
also be obtained easily. Before permuting back to the original order
and restoring the signs, $l-1$ alternative neighboring approximations
$\tilde{\mathbf{z}}$ can be obtained by adjusting the entries of
$\hat{\mathbf{z}}^{(N)}$ in the
following way: $\tilde{z}_{r}=\hat{z}_{r}^{(N)}+1$ if $\hat{z}_{r}^{(N)}%
=l_{r}$ and $\tilde{z}_{r}=\hat{z}_{r}^{(N)}-1$ if
$\hat{z}_{r}^{(N)}=l_{r}+1$.

\subsection{Search for $\hat{\alpha}_{r}(\theta)$}\label{app:iter}

For the first axis to be simplified, $\hat{\alpha}_{1}(\theta)$ is
approximated by the $N$-ratio simplification of $\mathbf{q}_{1}$ with
the smallest $N$ that satisfies the minimum accuracy required
$\cos(\theta)$. For $r\geq2$, the orthogonality restrictions need to be
taken into account. Here, we search for $\hat{\alpha }_{r}(\theta)$
using a hybrid approach which takes the best solution out of the three
different procedures described below (two in Section \ref{app:alg.conv} and
one in Section \ref{app:alg.orthog}), as ranked first by the smallest value of
$N_{r}(\theta)$ found, and then by accuracy.

We denote by $\mathbf{H}_{r-1}$ the matrix representing orthogonal
projection
onto $\mathcal{N}(\mathbf{Z}_{r-1}^{\top})$, the null space of $\mathbf{Z}%
_{r-1}^{\top}$, where $\mathbf{Z}_{r-1}$ is any $p\times(r-1)$ matrix
whose columns are integer representations of the axes already
simplified, $\hat{\alpha}_{1},\ldots,\hat{\alpha}_{r-1}$. As detailed
in Section \ref{app:hmat} below, $\mathbf{H}_{r-1}=\widetilde
{\mathbf{H}}_{r-1}/N_{r-1}$ for some known integer matrix $\widetilde
{\mathbf{H}}_{r-1} $ and positive integer $N_{r-1}$.

\subsubsection{Algorithms based on convergence to orthogonality}\label{app:alg.conv}

We describe here two versions of an iterative algorithm to find an axis
of minimal complexity that satisfies the orthogonality and minimum
accuracy restrictions. Starting with $N=1$, the algorithm works by
first obtaining the
$N$-ratio simplification of $\mathbf{q}_{r}^{\perp}=\mathbf{H}_{r-1}%
\mathbf{q}_{r}$ and then modifying it, directly controlling its
complexity, while aiming to maintain accuracy and improving the degree
to which the orthogonality conditions are met.

The algorithm is based on the function $0<\omega(\mathbf{z}):=\mathit{accu}(\mathbf{z}%
,\mathbf{H}_{r-1}\mathbf{z})\leq1$ which measures the closeness of
$\ell(\mathbf{z})$ to $\mathcal{N}(\mathbf{Z}_{r-1}^{\top})$, the
orthogonality conditions being met if and only if
$\omega(\mathbf{z})=1$.

The algorithm has three stages:

\begin{description}
\item[Stage 1.] [1] Compute $\hat{\mathbf{z}}^{(N)}$, the $N$-ratio
simplification of $\mathbf{q}_{r}^{\perp}$. [1$^{\ast}$] If $\omega
(\hat{\mathbf{z}}^{(N)})=1$ and $\hat{\mathbf{z}}^{(N)}$ satisfies the
minimum accuracy required, we take $\hat{\alpha}_{r}(\theta)$ to be
$\ell
(\hat{\mathbf{z}}^{(N)})$ and the algorithm stops. If $\omega(\hat{\mathbf{z}%
}^{(N)})=1$, but $\hat{\mathbf{z}}^{(N)}$ does not satisfy the minimum
accuracy required, we update $N\leftarrow N+1$ and return to [1].
Otherwise, $\omega(\hat{\mathbf{z}}^{(N)})<1$ and we move on to Stage
2.

\item[Stage 2.] Construct a set of neighbor vectors $\mathcal{Z}%
\subset\mathbb{Z}^{(p)}$ by increasing and decreasing one of the
entries of $\hat{\mathbf{z}}^{(N)}$ by one unit (see Section \ref{app:nratio}),
identifying its (possibly empty) subset $\mathcal{Z}_{1}$ of vectors
with $\omega(\mathbf{z})=1$. If there is a
$\mathbf{z}\in\mathcal{Z}_{1}$ satisfying the minimum accuracy
required, we take $\hat{\alpha}_{r}(\theta)$ to be the most accurate
such vector and the algorithm stops. If there is a
$\mathbf{z}\in\mathcal{Z}_{1}$, but no such vector satisfies the
minimum accuracy required, we update $N\leftarrow N+1$
and return to [1]. Otherwise, $\omega(\mathbf{z})<1$ for all $\mathbf{z}%
\in\mathcal{Z}$ and we identify its (possibly empty) subset $\mathcal{Z}%
(\theta)$ of vectors satisfying the minimum accuracy required. If
$\mathcal{Z}(\theta)$ is the empty set, we move on to Stage 3.
Otherwise, we
set $\mathbf{z}^{\prime}$ to be $\operatorname{\arg\,\max}\{\omega(\mathbf{z})\dvtx \mathbf{z}%
\in\mathcal{Z}(\theta)\}$. If $\omega(\mathbf{z}^{\prime})\leq\omega
(\hat{\mathbf{z}}^{(N)})$, we again move on to Stage 3. Otherwise, if
$\omega(\mathbf{z}^{\prime})>\omega(\hat{\mathbf{z}}^{(N)})$, we update
$\hat{\mathbf{z}}^{(N)}\leftarrow\mathbf{z}^{\ast}$ (defined below) and
return to [1$^{\ast}$]. We have two variants of this algorithm,
corresponding to two different choices of $\mathbf{z}^{\ast}$:

\begin{enumerate}
\item $\mathbf{z}^{\ast}=\mathbf{z}^{\prime}$: this hungrily pursues
orthogonality, at a potential loss of accuracy.\vspace*{1pt}

\item $\mathbf{z}^{\ast}=\mbox{arg}\max\{\mathit{accu}(\mathbf{q}_{r}^{\perp
},\mathbf{z}) \dvtx\omega(\mathbf{z})>\omega(\hat{\mathbf{z}}^{(N)}%
) , \mathbf{z}\in\mathcal{Z}(\theta)\}$: this retains accuracy as
much as possible, while improving the extent to which the orthogonality
conditions are met.
\end{enumerate}

\item[Stage 3.] We construct a set of higher order neighbor vectors
$\mathcal{Z}$ by moving more than one entry of $\hat{\mathbf{z}}^{(N)}$
in the
direction defined by the integer vector $\widetilde{\mathbf{H}}_{r-1}%
\hat{\mathbf{z}}^{(N)}-N_{r-1}\hat{\mathbf{z}}^{(N)}$. We then follow
the same procedure as in Stage 2 except that, if $\mathcal{Z}(\theta)$
is empty or
$\omega(\mathbf{z}^{\prime})\leq\omega(\hat{\mathbf{z}}^{(N)})$, we now
update $N\leftarrow N+1$ and return to [1].
\end{description}

\begin{remark} If we obtain a vector of complexity strictly bigger
than the current of $N$, we do not consider it at that stage, but keep
it for later feasibility, provided its complexity is not bigger than
$N^{\ast}$.
\end{remark}

\begin{remark} It is easy to show that, for any $\mathbf{z}\in
\mathbb{Z}^{(p)}$ with $\mathbf{H}_{r-1}\mathbf{z}\neq\mathbf{0}_{p}$,
\begin{eqnarray*}
\frac{\mathit{accu}(\mathbf{q}_{r}^{\perp},\mathbf{z})}{\mathit{accu}(\mathbf{q}_{r}^{\perp
},\mathbf{H}_{r-1}\mathbf{z})}&=&\frac{\Vert\mathbf{H}_{r-1}\mathbf{z}\Vert
}{\Vert\mathbf{z}\Vert}\\
&=&\mathit{accu}(\mathbf{z},\mathbf{H}_{r-1}\mathbf{z}%
)\leq1,%
\end{eqnarray*}
so that $\mathit{accu}(\mathbf{q}_{r}^{\perp},\mathbf{H}_{r-1}\mathbf{z})\geq
\mathit{accu}(\mathbf{q}_{r}^{\perp},\mathbf{z})$, equality holding if and only
if $\mathbf{z}$ obeys the orthogonality conditions
$\mathbf{z=H}_{r-1}\mathbf{z} $. For any other $\mathbf{z}$, projection
strictly increases accuracy. Given the general trade-off between
accuracy and simplicity, this suggests that projection tends to
increase complexity. Accordingly, there is a premium on algorithms,
such as the one just described, which avoid projection \textit{per se}.
\end{remark}

\subsubsection{Algorithm based on exact orthogonality}\label{app:alg.orthog}

The following algorithm ensures exact orthogonality at every step by
restricting attention to axes of the form
$\ell(\mathbf{O}_{r-1}\mathbf{y})$,
$\mathbf{y}\in\mathbb{Z}^{(p-r+1)}$, where $\mathbf{O}_{r-1}$ is a
$p\times(p-r+1)$ integer matrix whose columns form a basis of $\mathcal{N}%
(\mathbf{Z}_{r-1}^{\top})$. The particular matrix $\mathbf{O}_{r-1}$\
used, which appears to work well, mitigates the fact that the
complexity and accuracy of $\mathbf{z=O}_{r-1}\mathbf{y}$ are
indirectly controlled; see Section \ref{app:orthog.basis}.

Putting $\mathbf{y}^{\ast}:=(\mathbf{O}_{r-1}^{\top}\mathbf{O}_{r-1}%
)^{-1}\mathbf{O}_{r-1}^{\top}\mathbf{q}_{r}$, $\mathbf{O}_{r-1}\mathbf{y}%
^{\ast}=\mathbf{q}_{r}^{\perp}$ is the closest point to
$\mathbf{q}_{r}$ in
$\mathcal{N}(\mathbf{Z}_{r-1}^{\top})$. Whereas the elements of $\mathbf{y}%
^{\ast}$ will not in general be integers, we may obtain an
approximation $\tilde{\alpha}_{r}(\theta)$ to
$\hat{\alpha}_{r}(\theta)$ as follows:

\begin{enumerate}
\item Compute the set of integer vectors $\mathcal{Y}\subset\mathbb{Z}%
^{(p-r+1)}$ obtained by $N$-ratio simplification of
$\mathbf{y}^{\ast}$, together with their angle neighbors, for all
$N\leq N^{\ast}$.

\item Obtain the set $\ell(\mathbf{O}_{r-1}\mathcal{Y})$ of all axes
$\ell(\mathbf{O}_{r-1}\mathbf{y})$ with $\mathbf{y}\in\mathcal{Y}$, and
find the minimum complexity $\tilde{N}_{r}(\theta)$ over all axes in
this set which satisfy the minimum accuracy requirement $\cos(\theta)$.

\item Call $\tilde{\alpha}_{r}(\theta)$ the most accurate axis in
$\ell(\mathbf{O}_{r-1}\mathcal{Y})$ with complexity
$\tilde{N}_{r}(\theta)$.
\end{enumerate}

\subsubsection{Choice of $\mathbf{O}_{r-1}$} \label{app:orthog.basis}

The choice $\mathbf{O}_{0}=\mathbf{I}_{p}$ is clearly optimal. For
$r>1$, the choice of $\mathbf{O}_{r-1}$ depends on an initial
permutation of the rows of $\mathbf{Z}_{r-1}$---defined below---such
that the first $r-1$ are linearly independent, forming a nonsingular
matrix $\mathbf{Z}_{a}$ in the corresponding partition
$\mathbf{Z}_{r-1}^{\top}=(\mathbf{Z}_{a}^{\top
}  \mathbf{Z}_{b}^{\top})$. This permutation is inverted at the end
to maintain the identity of the variables.

Conformably partitioning $\mathbf{u}\in\mathbb{R}^{p}$ as
$\mathbf{u}^{\top }=(\mathbf{u}_{a}^{\top}  \mathbf{u}_{b}^{\top})$,
$\mathbf{u}\in
\mathcal{N}(\mathbf{Z}_{r-1}^{\top})$ when $\mathbf{Z}_{a}^{\top}%
\mathbf{u}_{a}+\mathbf{Z}_{b}^{\top}\mathbf{u}_{b}=(0,\ldots,0)^{\top}$.
Equivalently, $\det(\mathbf{Z}_{a}) \mathbf{u}_{a}%
=-\operatorname{cof}(\mathbf{Z}_{a})\mathbf{Z}_{b}^{\top}\mathbf{u}_{b}$, where
cof$(\mathbf{Z}_{a})$ is the matrix of cofactors of $\mathbf{Z}_{a}$.
Thus,
\[
\mathbf{O}_{r-1}:=\pmatrix{
-\operatorname{cof}(\mathbf{Z}_{a})\mathbf{Z}_{b}^{\top}\cr
\det(\mathbf{Z}_{a}) \mathbf{I}_{p-r+1}%
}
\]
is an integer matrix whose columns form a basis of $\mathcal{N}(\mathbf{Z}%
_{r-1}^{\top})$.

For any $\mathbf{y}\in\mathbb{Z}^{(p-r+1)}$, conformably partitioning
$\mathbf{z=O}_{r-1}\mathbf{y}$ as
$\mathbf{z}^{\top}=(\mathbf{z}_{a}^{\top }  \mathbf{z}_{b}^{\top})$
gives $\ell(\mathbf{z}_{b})=\ell(\mathbf{y})$. We choose the initial
permutation of the rows of $\mathbf{Z}_{r-1}$ so that the elements of
$\mathbf{q}_{r}^{\perp}$ corresponding to $\mathbf{z}_{b}$ have the
largest possible set of absolute values, these contributing most to
angle-accuracy. Specifically, we proceed as follows. First, permute the
elements of $\mathbf{q}_{r}^{\perp}$ so that the absolute values of
its elements are in increasing order, permuting the rows of
$\mathbf{Z}_{r-1} $ accordingly. Find the first set of $r-1$ rows of
$\mathbf{Z}_{r-1}$ having nonzero determinant in the lexicographical
ordering of such sets by their row labels. Finally, maintaining the
internal ordering of these rows (and of their
$p-r+1$ complementary rows), make them the first $r-1$ rows, $\mathbf{Z}_{a}%
$, of a new matrix $\mathbf{Z}_{r-1}$.

\subsubsection{Construction of the projector $\mathbf{H}_{r-1}$}\label{app:hmat}

The matrix $\mathbf{H}_{r-1}$ is proportional to an integer matrix, so
that $\mathbf{H}_{r-1}=\widetilde{\mathbf{H}}_{r-1}/N_{r-1}$ for some
integer matrix $\widetilde{\mathbf{H}}_{r-1}$ and positive integer
$N_{r-1}$. Simple updates are available to construct this matrix.

Putting $N_{0}=1$ and
$\mathbf{H}_{0}=\widetilde{\mathbf{H}}_{0}=I_{p}$, for
each $r\geq1$, $\mathbf{H}_{r}=\mathbf{H}_{r-1}-\hat{\mathbf{z}}_{r}%
\hat{\mathbf{z}}_{r}^{\top}/\Vert\hat{\mathbf{z}}_{r}\Vert^{2},$ so
that
$\mathbf{H}_{r}=\widetilde{\mathbf{H}}_{r}/N_{r}$ with $\widetilde{\mathbf{H}%
}_{r}=[\Vert\hat{\mathbf{z}}_{r}\Vert^{2}\widetilde{\mathbf{H}}_{r-1}%
-N_{r-1}\hat{\mathbf{z}}_{r}\hat{\mathbf{z}}_{r}^{\top}]/h_{r}$ and
$N_{r}=[N_{r-1},\times\break\Vert\hat{\mathbf{z}}_{r}\Vert^{2}]/h_{r}$, in which
$h_{r}=\operatorname{hcf}(N_{r-1}\Vert\hat{\mathbf{z}}_{r}\Vert^{2})$. The
simplicity of these updates is another advantage of requiring
orthogonality.
\end{appendix}

\section*{Acknowledgments}

We are grateful to Paddy Farrington and Chris Jones for useful comments
on earlier versions of this manuscript, and to Nickolay Trendafilov for
helpful discussions.

\printaddresses

\end{document}